 \documentclass[final,5p,times,twocolumn]{elsarticle}

\usepackage{amssymb}
\usepackage{lipsum}
\usepackage[fleqn]{amsmath}
\newcommand{\braket}[2]{\left \langle #1 \middle| #2 \right \rangle}
\newcommand{\braketmatrix}[3]{\left \langle #1 \middle| #2 \middle| #3 \right \rangle}

\usepackage[inline]{enumitem}

\journal{Physica D: Nonlinear Phenomena}

\begin{document}

\begin{frontmatter}



\title{Targeted energy transfer dynamics and chemical reactions}

\author{N. Almazova} 
\author{S. Aubry}%
\author{G. P. Tsironis}
\address{Institute of Theoretical and Computational Physics and Department of Physics, University of Crete, Heraklion, 71003, Greece}

\begin{abstract}
  Ultrafast reaction processes take place when resonant features of nonlinear model systems are taken into account.  In the targeted energy or electron transfer dimer model this is accomplished through the implementation of nonlinear oscillators with opposing types of nonlinearities, one attractive while the second repulsive.  In the present work we show that this resonant behavior survives if we take into account the vibrational degrees of freedom as well.  After giving a summary on the basic formalism of chemical reactions we show that resonant electron transfer can be assisted by vibrations.  We find the condition for this efficient transfer and show that in the case of additional interaction with noise a distinct non-Arrhenius behavior develops that is markedly different from the usual Kramers-like activated transfer.
\end{abstract}

\end{frontmatter}

\section{Introduction}
The Targeted Energy Transfer (TET) model considers two weakly coupled nonlinear oscillators that have soft and hard nonlinearity respectively and finds a parameter regime where perfect resonant transfer occurs between the oscillators \cite{kopidakis2001targeted, aubry2001analytic}. There is by now extended literature on this model, both in the classical, quantum but also in the engineering regime with a more practical application of the concept \cite{vakakis2008nonlinear}. Recently, a machine learning approach was used in both classical and quantum regimes that explores the possibility of finding the precise transfer resonance through learning processes \cite{barmparis2021discovering, andronis2023quantum}. This method is quite promising since it in principle enables the extension of the TET concept to arbitrary size systems, a feat that cannot easily accomplished through more traditional numerical techniques.

The concept of targeted transfer is motivated from biology and in particular by the specificity and efficiency that certain electron transfer processes have in biological systems\cite{aubry2005nonadiabatic}. In chlorophyll, in particular, there is a transfer electronic path that is ultra fast and, as a result, it is bound to involve certain resonant transfer features. The TET mechanism provides a simplified framework for this ultra-fast transfer through the generation of a specific transfer between non-identical nonlinear oscillators that is practically perfect. 

One feature that is missing in the original TET formulation if one wants to address more realistically the electron transfer processes is the presence and interaction with additional degrees of freedom and specifically phonons. While this is done in an indirect way through the use of the Discrete Nonlinear Schr{\"o}dinger (DNLS) Equation model, for a more complete and realistic analysis one needs to include explicitly these degrees of freedom. This is accomplished through the use of a model that includes both electronic and vibrational degrees of freedom as well as their coupling. The main target of the present work is to focus precisely on this microscopic case, viz. a complete electron-phonon model, and address the TET condition in this more general and mode realistic case. This study involves a deeper understanding of the chemical reaction processes and aspects that may affect their efficiency. In order to explore the complete power of the chemistry under TET we provide first a description of chemical reactions and explore the differences of TET driven reactions compared to more standard ones. 

The plan of this paper is then the following: In the next section we discuss chemical reactions and introduce the diabatic approximation for their description. We describe the Born-Oppenheimer framework and detail the Markus theory of chemical reactions. We work explicitly with a two state model and present the reaction dynamics in the semiclassical approximation. We use adiabatic surfaces in order to discuss conical intersections. Subsequently, we introduced targeted electron/energy transfer through a simplified two state model that also includes semiclassically vibrational degrees of freedom. We show that this model has radically different behavior compared to the standard Markus law and investigate the temperature dependence of the transfer process. In the presence of noise, we observe a clear non-Arrhenius behavior that is quite distinct from the typical of standard chemical reactions. In the concluding section, we summarize our findings and give an outlook for the generalization of this work.
\section{\label{sec1} Chemical Reactions}
Formulas are widely used in chemistry for describing the organisation of molecules, radicals, and complexes \cite{goodwin2008structural,brown2009chemistry,batsanov2012introduction,michl2003organic}. In their most detailed forms, i.e. the "condensed formulas", they describe schematically how the nuclei constituents of a chemical species are spatially organized and bounded by covalent single, double, triple bonds, hydrogen bonds, Van de Waals bonds, etc.
They also specify where the charged atoms or radicals are located, and the pending bonds, etc. while they also give spatial information about the organization of a single molecule or an aggregate of molecules.
Thus, the condensed formula of a molecule or radical is nothing but a characterization of its electronic state as precisely as possible.

Chemical reactions are processes in which one or more substances, known as the reactants, are chemically transformed into one or more new substances, called the products \cite{march1977advanced, lagana2018chemical}. These changes involve the creation or the breaking of chemical bonds or a charge transfer that generates relative nuclei displacements and thus molecular reorganization. These changes are schematically represented by a symbolic reaction formula which describes the changes that occur in the the substances.
In many cases, they consist of a collection of subsequent elementary processes, referred to as elementary steps or elementary reactions, that describe how the overall reaction proceeds.

We focus essentially on Elementary Chemical Reactions (ECR) introduced through the Diabatic representation. Diabatic states are defined empirically as an implicit consequence of the theory of chemical bonds pioneered by L. Pauling \cite{pauling1931nature}. They are based on orbital occupation, non-occupation, and related overlaps.
This theory is remarkably successful because it provides the basis for the concept of the chemical formula and the valence rules which suffer only rare exceptions, such as for the intermediate valence materials.

Next, we discuss the validity of the Diabatic approximation within the Born Oppenheimer representation which has well defined foundations. It is also widely used in quantum chemistry although its flaw is that it does not suggest any intuitive representation of the molecular arrangements and their chemical formula.

\subsection{Elementary chemical reactions in the diabatic representation}
The following exposition draws on a more extended earlier report \cite{aubry2014semiclassical}. The work has been done to describe long-range inter-molecular photo-initiated electron transfer using Potential Energy Surfaces (PES) \cite{worth2001mediation, farfan2020systematic}. Here we use the semiclassical approximation from the beginning, in other words, we describe the dynamics of the nuclei by classical variables instead of quantum operators. 
The electrons still remain quantum, however, we project their wave function into a two dimensional subspace in order to deal with only two complex variables with the unit norm.

An ECR is defined as a chemical transformation process whereby an initial state, described by a specific chemical formula, is converted into a final state, represented by a different chemical formula.
We consider two species, that can be reacting molecules, radicals, or clusters that are designated each with the chemical formula $D$ and $A$ respectively and that are supposed to exchange, for instance, an electron; this process is represented by the chemical reaction $D + A \rightarrow D^{+}A^{-}$.

Subsequently, we assume the existence of two diabatic states $\chi_D(\{r_{\nu}\};\{R_n\})$ and $\chi_A(\{r_{\nu}\};\{R_n\})$ corresponding to the initial and the final states, respectively.
More precisely, the quantity $\chi_D(\{r_{\nu}\};\{R_n\})$ represents the real global electronic wave function of the set of electronic variables $\{r_{\nu}\}$ for the system in the initial state $D+A$. This wave function also depends on the nuclei coordinates $\{R_n\}$ supposed to be classical variables and therefore appear as parameters. The second diabatic state $\chi_A(\{r_{\nu}\};\{R_n\})$ is defined identically but for the global system in the final state $ D^+ A^{-}$. These two wave functions a priori are not orthogonal. It is assumed that they can be orthogonalized with slight perturbations. We thus assume for all nuclei coordinates $\{R_n\}$ $\braket{\chi_D(\{r_{\nu}\}; \{R_n\})}{\chi_A(\{r_{\nu}\}; \{R_n\})}_e =0$, where the index $e$ specifies that the scalar product involves integration only over the electronic degrees of freedom. The nondiabatic theory of ET described in \cite{aubry2005nonadiabatic}

Since we study the transition between the initial state and the final state, we suggest that during that transition the electronic wave function has the form
\begin{equation}
    \chi(\{r_{\nu}\};\{R_n\}) =\varphi_D(t)\chi_D(\{r_{\nu}\};\{R_n\}) + \varphi_A(t) \chi_A(\{r_{\nu}\}; \{R_n\})
    \label{diabrep}
\end{equation}
where $\varphi_D(t) $ and $\varphi_A(t)$ are time dependant complex coefficients which fulfill the normalization condition $|\varphi_D(t)|^2+ |\varphi_A(t)|^2 =1$.

The global quantum Hamiltonian of the system can be expressed as follows
\begin{equation}
    \mathbf{H}=\mathcal{H}_e(\{r_{\nu}\},\{p_{\nu}\};\{R_n\})+\mathbf{H}_K
    \label{gH0}
\end{equation}
The term $\mathbf{H}_e(\{r_{\nu}\},\{p_{\nu}\};\{R_n\})$ denotes the portion of the Hamiltonian that concerns exclusively to the electrons $\nu$ considered as fermions. The kinetic energy, given by $1/(2 m_e)\: p^2_{\nu}$ for all the electrons (with mass $m_e$) functions of the momentum operators ($p_{\nu}$) and all potential Coulomb interactions, including those between electrons, electrons and nuclei, and nuclei themselves.
$\mathcal{H}_K= \sum_n 1/(2M_n)\: P_n^2 $ is the total kinetic energy operator of the nuclei with coordinates $R_n$, conjugate variables $P_n$ and mass $M_n$. Then we can define electronic energy.
\begin{align}
\begin{split}
    \mathbf{V} (\varphi_D,& \varphi_A, \{R_n\}) = \\
    &=\braketmatrix{\chi(\{r_{\nu}\}; \{R_n\})}{\mathcal{H}_e(\{r_{\nu}\},\{p_{\nu}\};\{R_n\})}{\chi(\{r_{\nu}\}; \{R_n\})}_e \\
    &= \begin{pmatrix} \varphi_D & \varphi_A \end{pmatrix} \cdot 
    \begin{pmatrix} E_D( \{R_n\}) & \Lambda( \{R_n\})\\ \Lambda( \{R_n\}) & E_A( \{R_n\}) \end{pmatrix} \begin{pmatrix} \varphi_D \\\varphi_A \end{pmatrix}
\end{split}\label{matform}
\end{align}
where
\begin{align}
    &E_D( \{R_n\}) = \braketmatrix{\chi_D}{\mathcal{H}_e}{\chi_D}_e \\
    &E_A( \{R_n\}) = \braketmatrix{\chi_A}{\mathcal{H}_e}{\chi_A}_e \\
    &\Lambda( \{R_n\}) = \braketmatrix{\chi_D}{\mathcal{H}_e}{\chi_A}_e=\braketmatrix{\chi_A}{\mathcal{H}_e}{\chi_D}_e
\end{align}
are real functions of the nuclei coordinates.

We can obtain an effective classical Hamiltonian written as
\begin{equation}
    \mathbf{H}_{diab} = \mathbf{H}_K + \mathbf{V} (\varphi_D,\varphi_A, \{R_n\})
    \label{hamdiab}
\end{equation}

The Hamiltonian in Eq. (\ref{hamdiab}) describes the dynamics of the electronic state projected in the two dimensional subspace defined by Eq. (\ref{diabrep}) coupled now with the set of nuclei coordinates.
The corresponding Hamilton equations for the nuclei and the electronic state are
\begin{align}
    i\hbar \dot{\varphi}_D &= E_D( \{R_n\}) \varphi_D+ \Lambda( \{R_n\}) \varphi_A  \label{hameq1}\\
    i\hbar \dot{\varphi}_A &= \Lambda( \{R_n\}) \varphi_D + E_A( \{R_n\}) \varphi_A   \label{hameq2}\\
    \begin{split}
        M_n \ddot{R}_n  &+ \frac{\partial E_D}{\partial R_n} |\varphi_D|^2 +\frac{\partial E_A}{\partial R_n} |\varphi_A|^2 \\
    &\qquad \qquad\quad+\frac{\partial \Lambda}{\partial R_n} (\varphi_D^{\star} \varphi_A+\varphi_A^{\star} \varphi_D) = 0 \label{hameq3}
    \end{split}
\end{align}

It is convenient to rewrite this Hamiltonian Eq. (\ref{hamdiab}) as a spin boson system. For that we expand  the $2\times 2$ in Eq. (\ref{matform}) on the base Pauli matrices
\begin{align}
    \begin{split}
        &\sigma_0 = \begin{pmatrix} 1 & 0 \\ 0 & 1 \end{pmatrix} \\
        &\sigma_x = \begin{pmatrix} 0 & 1 \\ 1 & 0 \end{pmatrix}, \quad
        \sigma_y = \begin{pmatrix} 0 & -i\\ i & 0 \end{pmatrix}, \quad
        \sigma_z = \begin{pmatrix} 1 & 0 \\ 0 &-1 \end{pmatrix}, \quad
    \end{split}\label{pauli_matrix}
\end{align}
which yields a Hamiltonian for nuclei coupled anharmonically to a spin operator, where $\varphi_D \vert{\uparrow}\rangle +\varphi_A \vert{\uparrow}\rangle$ represents the general state including the spin. This is expressed as follows:
\begin{align}
    \mathbf{H}_{diab} = \mathbf{H}_K + V_0 (\{R_n\})+h_x(\{R_n\}) \sigma_x + h_z(\{R_n\}) \sigma_z
    \label{hamdiabsb}
\end{align}
where
\begin{align}
    V_0 (\{R_n\}) &= \frac{1}{2} (E_D(\{R_n\})+E_A(\{R_n\}) \label{potential}\\
    h_x(\{R_n\}) &= \Lambda( \{R_n\})  \label{covalent}\\
    h_z(\{R_n\}) &=\frac{1}{2} (E_D(\{R_n\})-E_A(\{R_n\}) \label{ionic}
\end{align}
    
We note that the $z$-coupling terms $ h_z(\{R_n\})$ in the Hamiltonian Eq. (\ref{hamdiabsb}) favors ionicity, i.e. an eigenvector of the spin component $\sigma_z$ which are $\vert{\uparrow}\rangle$ or $\vert{\downarrow}\rangle$ (corresponding to the initial diabatic state and the final diabatic state).
On the other hand, the $x$-coupling term $ h_x(\{R_n\})$ favors covalence, i.e. an eigenvector of the spin component $\sigma_x$, which are ${1/\sqrt{2}}(\vert{\uparrow}\rangle +\vert{\downarrow}\rangle)$ or ${1/\sqrt{2}}(\vert{\uparrow}\rangle -\vert{\downarrow}\rangle)$ transverse to the $z$ component.
In the context of electron transfer theory, the transverse term is typically regarded as a relatively minor contributor.

\subsection{ The Marcus Theory }
We consider the example of an elementary reaction which consists on an electron transfer between a donor and an acceptor. The outer sphere electron transfer (OET) has been studied intensively by R. Marcus \cite{marcus1992rudolph, RevModPhys65599}. The reaction process may be interpreted as the thermally activated jump of a single electron from an orbital near a donor site $D^{-}$ to another orbital near an acceptor site $A$ symbolically represented by the chemical reaction $D^{-} A \rightarrow D A^{-}$ \cite{marcus1985electron}.
The Marcus model for OET is based on the (empirical) Diabatic representation of elementary chemical reactions. It assumes that the covalent interactions favoring hybridization between the donor and acceptor are small compared to the ionic interactions favoring the localization of the electron either on the donor or the acceptor sites. This covalent interaction plays a possible role only near the transition state where it can favor the jump of the electron
between two potential energy surfaces. There are other forms of electron transfer (ET) called inner sphere electron transfer (IET) with more complex descriptions involving transient covalent binding.

\begin{figure}[ht]\centering
    \includegraphics[width=0.8\columnwidth]{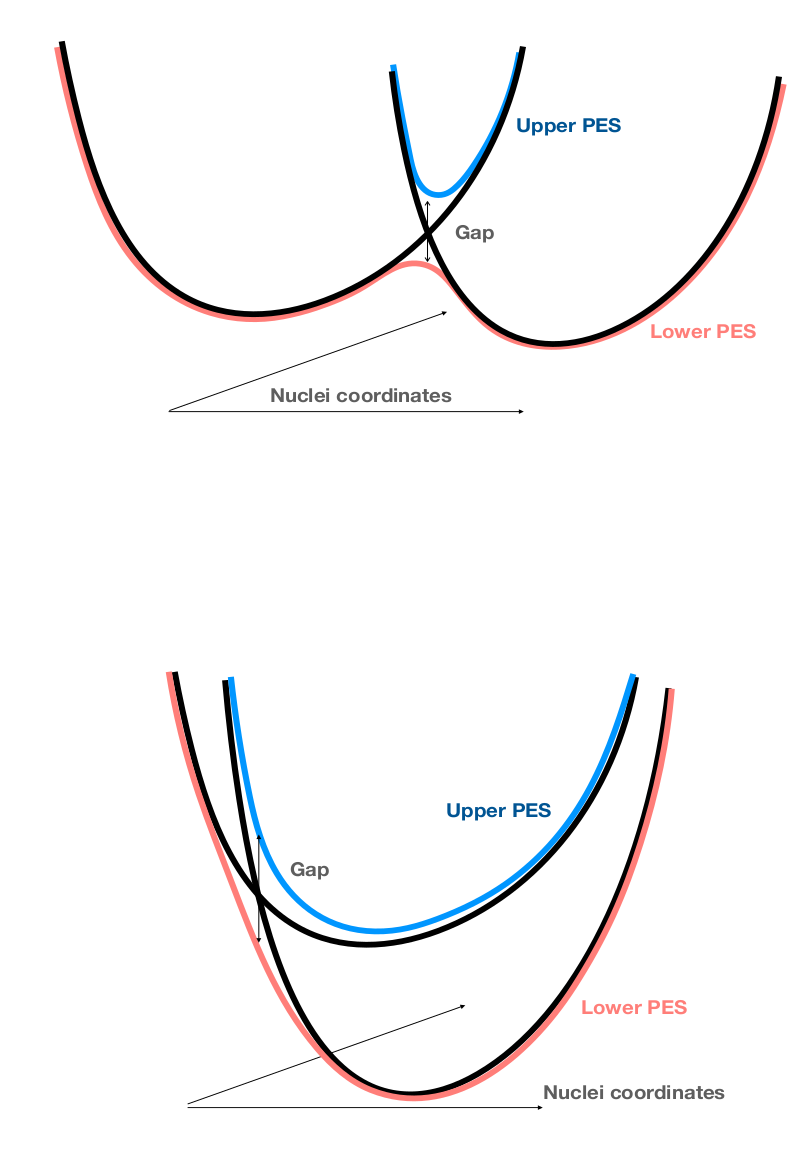}
    \caption{ Two schemes showing two Diabatic energy surfaces (black curves) representing the initial state before reaction and the final state after reaction. Opening a gap at the intersection of these surfaces generates two PES   (red and blue curves respectively). The upper scheme corresponds to the Marcus scheme of ET in the normal regime while the lower scheme corresponds to the inverted regime.} \label{fig:pes_general_marcus}
\end{figure}

\begin{figure}[ht]\centering
    \includegraphics[width=0.8\columnwidth]{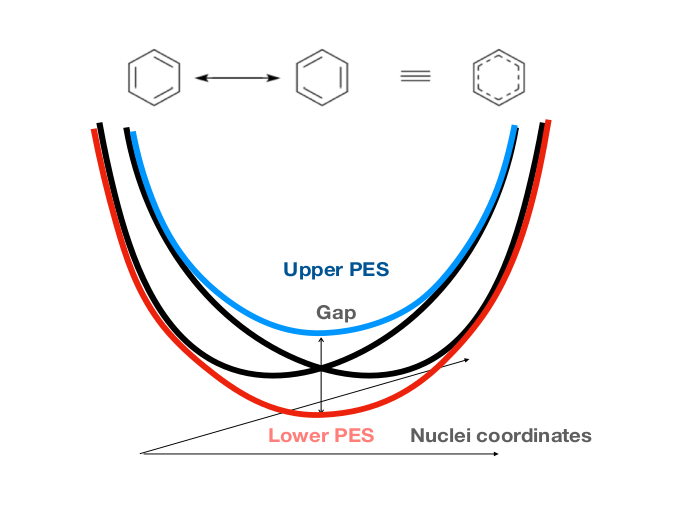}
    \caption{Another scheme showing two Diabatic energy surfaces (black curves) and the two resulting PES (red and blue curve respectively) with a large gap opening. This situation is not associated with any chemical reaction.
        The diabatic states may correspond for instance to the two possible chemical formulas of Benzene while the PES corresponds to the real state which is a well-known hybridized state between the two diabatic states (which can be symmetric or antisymmetric)
        .} \label{fig:pes_large_gap}
\end{figure}

Although the diabatic states may look rather well defined far enough from the intersection between the two diabatic energy surfaces, they are only empirically defined especially in the vicinity of their intersection. Despite the fact that it is known that generally the diabatic representation cannot be defined rigorously {\cite{smith1969diabatic,baer1975adiabatic,baer2006beyond}, it is nevertheless commonly used in quantum chemistry. The main possibilities considered for the donor-acceptor system are illustrated in Figs. \ref{fig:pes_general_marcus} and \ref{fig:pes_large_gap}. In Fig. \ref{fig:pes_general_marcus} two intersecting surfaces lead to the so-called normal (upper) and inverted (lower) Markus regimes. An energy gap opens up in these two cases, which leads to specific Arrhenius-type exchanges. In Fig. \ref{fig:pes_large_gap}, we consider another energy surface intersection possibility that is not connected to a specific chemical reaction. In the following subsection, we discuss how the diabatic curves can be related to the Potential Energy Surfaces (PES) defined within the Born Oppenheimer representation.

An extension of the original Marcus model was already introduced where the assumption was that the covalent interactions are stronger and comparable to the ionic interactions and an interesting intermediate regime was discovered where coherent Ultrafast Electron Transfer (UET) may occur \cite{aubry2014semiclassical}.

\section{The  Born Oppenheimer Representation}
The Born Oppenheimer (BO) representation has a rigorous definition that does not explicitly include terms that lead to chemical formulas. 
It was initially developed for a non-relativistic model consisting of a finite collection of electrons (fermions)  and nuclei considered spinless particles where magnetic interactions are neglected.
The whole system obeys a Schr\"odinger equation with a complex global Hamiltonian $\mathbf{H}$.
It is always possible to split it as the sum of two parts
\begin{equation}
    \mathbf{H}=\mathcal{H}_e(\{r_{\nu}\},\{p_{\nu}\};\{R_n\})+\mathcal{H}_K
\end{equation}\label{gH}
where the kinetic energy operator of the nuclei
\begin{equation}
    \mathcal{H}_K= \sum_n \frac{1}{2M_n} P_n^2
    \label{kin}
\end{equation}
is separated from the rest of the Hamiltonian; the latter is called the electronic part of the Hamiltonian.
\begin{equation}
    P_n= \frac{\hbar}{i} \frac{\partial.}{\partial R_n}
    \label{conjv}
\end{equation}
denote the conjugate quantum operator associated with the nuclei variable $R_n$.

The electronic Hamiltonian $\mathcal{H}_e(\{r_{\nu}\},\{p_{\nu}\};\{R_n\})$ is the sum of the kinetic energy operators of the electrons and of all the Coulomb interactions potential in the system that is 
\begin{enumerate*}
        \item the electrons one with each other,\item the electrons with the nuclei, \item between the nuclei with each other. (which is generated only by the Coulomb potential of the nuclei with coordinates $\{R_n\}$ and does not affect the $\{r_{\nu}\}$ and their conjugate operators $\{p_{\nu}\}$).
\end{enumerate*}

As commented previously, the possible interactions due to the magnetic fields which could be generated by the spin of the electrons and/or their orbitals are neglected here as well as many BO theories.
If one neglects the nuclei kinetic energy $\mathcal{H}_K $ the nuclei coordinates $\{R_n\}$ appear as real parameters in $\mathcal{H}_e( \{R_n\})$. 
We may then proceed to diagonalize this electronic component formally, namely $\mathcal{H}_e(\{r_{\nu}\},\{p_{\nu}\};\{R_n\})$. The sequence of electronic levels, $E_i(\{R_n\})$, is defined in increasing order with $i=0,1,... +\infty$ associated with electronic eigenfunctions $\chi_i(\{r_{\nu}\}; \{R_n\})$ which are normalized and have to be antisymmetric under electron exchange (with the same spin) since the electrons are fermions.
These eigenfunctions are real since no magnetic interaction is involved in the electronic Hamiltonian Eq. (\ref{gH}).
The smallest electronic energy $E_0(\{R_n\})$ corresponds to the electronic ground state.

Then we can expand the electronic Hamiltonian on its base of eigenstates
\begin{align}\label{felH}
\begin{split}
    \mathcal{H}_e(\{r_{\nu}\},&\{p_{\nu}\};\{R_n\})= \\
    &= \sum_i E_i(\{R_n\})\cdot
    \vert \chi_i(\{r_{\nu}\}; \{R_n\})\rangle \langle\chi_i(\{r_{\nu}\}; \{R_n\})\vert
\end{split}
\end{align}
Similarly, the global wave function $\Phi(\{R_n\},\{r_{\nu}\})$ of the system described by Hamiltonian Eq. (\ref{gH}) can be also expanded in this electronic base $\vert\chi_i(\{r_{\nu}\}; \{R_n\})\rangle$ as
\begin{align}
    \Phi(\{R_n\}, \{r_{\nu}\}) = \sum_i \varphi_i(\{R_n\}) \chi_i(\{r_{\nu}\};\{R_n\})
    \label{formexp}
\end{align}
which defines a vector of wave functions $\bar{\varphi}(\{R_n\}) = \{\varphi_i(\{R_n\})\}$ for the nuclei variables.

If we eliminate the electronic variables $\{x_{\nu}\}$ the initial global Hamiltonian Eq. (\ref{gH}) becomes formally a new Hamiltonian acting on vectorial wave functions $\bar{\varphi}(\{R_n\})$.
This Hamiltonian takes the form of a matrix $\mathbf{H}(\{R_n\},\{P_n\})$ of operators $\hat{\mathbf{H}}_{i,j}(\{R_n\},\{P_n\})$
\begin{align}\label{opdef}
    \braketmatrix{\bar{\varphi}(\{R_n\})}{\mathbf{H}}{\bar{\varphi}(\{R_n\})} =\sum_{i,j} \braketmatrix{\varphi_i(\{R_n\})}{\hat{\mathbf{H}}_{i,j}}{\varphi_j(\{R_n\})}
\end{align}
where the matrix elements are defined as
\begin{align}
    \hat{\mathbf{H}}_{i,j} = E_i(\{R_n\}) \delta_{i,j} +\braketmatrix{\chi_i(\{r_{\nu}\}; \{R_n\})}{\mathcal{H}_K}{\chi_j(\{r_{\nu}\}; \{R_n\})}_e
    \label{matel}
\end{align}

The hermitian product $\langle.\vert.\rangle_e$ in this equation only involves integration over the electronic variables.
We calculate first $P_n\cdot \chi_i$ which is involved in the calculation of $\braketmatrix{\chi_i}{\mathcal{H}_K}{\chi_j}$. Then $P_n\cdot\chi_i= \frac{i}{\hbar}\frac{\partial }{\partial R_n}\chi_i(\{x_{\nu}\};\{R_n\})$ considered as a function of the electronic variables $\{x_{\nu}\}$ can be expanded on the base of electronic wave functions $\chi_j$ where $\{R_n\}$ are considered as parameters
\begin{align}\label{expans}
    P_n\cdot\chi_i(\{x_{\nu}\};\{R_n\}) =\frac{\hbar}{i} \sum_j b_{i,j}^{(n)}(\{R_n\})\chi_j(\{x_{\nu};\{R_n\}) 
\end{align}
Coefficients $b_{i,j}^{(n)} (\{R_n\}) $ are defined as
\begin{align}
     b_{i,j}^{(n)}(\{R_n\})= \braket{\chi_j(\{R_n\})}{\frac{\partial}{ \partial R_n} \chi_i(\{R_n\}) }_e
\end{align}\label{coeff}
These coefficients are real and obtained from standard perturbation theory for $i\neq j$
\begin{align}
    b_{i,j}^{(n)}(\{R_n\})= \frac{\braketmatrix{\chi_j(\{R_n\})}{\frac{\partial\mathcal{H}_e}{\partial R_n}}{\chi_i(\{R_n\})}_e}{E_i-E_j}
    \label{pcoeff}
\end{align}
while for $i=j$, the normalization of $\chi_i$ implies $b_{i,i}^{(n)}(\{R_n\})=0$. We note that $b_{i,j}^{(n)}\{R_n\})= - b_{j,i}^{(n)}(\{R_n\})$. Taking into account that variable $R_n$ and its conjugate variable $P_n$ considered as quantum operators do not commute one with each other, we obtain
\begin{align}
\begin{split}
    &\hat{\mathbf{H}}_{i,i}= \mathcal{H}_K +V_i(\{R_n\}) \quad \mbox{with}  \\
    &V_i(\{R_n\}) = E_i(\{R_n\}) -\sum_n  \frac{\hbar^2}{2M_n} \sum_k  b_{i,k}^{(n)} (\{R_n\})  b_{k,i}^{(n)} (\{R_n\}) \\
\end{split}
\label{Diag}
\end{align}
and for $i\neq j$
\begin{align}
    \begin{split}
    &\hat{\mathbf{H}}_{i,j} = C_{i,j} (\{R_n\}) + i D_{i,j} (\{R_n\})  \quad \mbox{with} \\
    &C_{i,j} (\{R_n\}) = - \sum_n \frac{\hbar^2}{2M_n} \sum_k  b_{i,k}^{(n)} (\{R_n\}) b_{k,j}^{(n)} (\{R_n\})\\
    &D_{i,j} (\{R_n\},\{P_n\}) =\sum_n \frac{\hbar }{2M_n} ( b_{i,j}^{(n)}(\{R_n\}) P_n + P_n b_{i,j}^{(n)}(\{R_n\})\\
    \end{split}\label{offdiag}
\end{align}

Finally, the dynamics of the nuclei coupled to the electrons take the usual simple form
\begin{equation}
    i \hbar \dot{\bar{\varphi}} =\hat{\mathbf{H}}\cdot\bar{\varphi}
    \label{scheq}
\end{equation}
where $\bar{\varphi}$ is a multicomponent wave functions defined by Eq.\ref{formexp}. There are no approximations to the initial global Hamiltonian (\ref{gH}) at this stage.

\subsection{ The Born-Oppenheimer approximation}
The Born-Oppenheimer (BO) approximation assumes that all off-diagonal terms $\hat{\mathbf{H}}_{i,j}$ ($i\neq j$) of the matrix $\mathbf{H}(\{R_n\},\{P_n\})$ of operators are negligible, such that the operator $\hat{\mathbf{H}}$ becomes diagonal. Consequently, we obtain a collection of independent BO Hamiltonians associated with each electronic eigenstate $i$:
\begin{align}
    \mathcal{H}_{BO} = \mathcal{H}_K + V_i(\{R_n\}) \label{BO_equation}
\end{align}
where the effective nuclei potentials are defined as $ V_i(\{R_n\}) = E_i(\{R_n\})+\Delta_{i,i}(\{R_n\})$.
Subsequently, the global wave function of the system when the system is situated on the $i^{th}$ PES can be expressed in a simple BO form
\begin{align}
    \Phi_i(\{R_n\}, \{r_{\nu}\}) = \varphi_i(\{R_n\}) \chi_i(\{r_{\nu}\};\{R_n\}) \label{appBO}
\end{align}

The electronic eigen-energy $i$ is defined in terms of the nuclei coordinates as a function of a potential energy surface (PES), which is denoted by the symbol $V_i(\{R_n\})$. The PES  are commonly observed in spectroscopy experiments and are studied in physics and chemistry.
Additionally, their values have been calculated numerically for simple molecules using ab initio methods or for the electronic ground state, $i=0$, employing Density Functional Theory (DFT). 

The minimum of the lowest PES $V_0(\{R_n\})$ yields the spatial arrangement of the nuclei system in its ground state. When approximating $V_0(\{R_n\})$ near its minimum by a harmonic expansion, the BO Hamiltonian can be diagonalized, thereby yielding the phonon spectrum and their associated modes possibly observable. The same procedure can be applied to the electronically excited states corresponding to different PES which yield different spatial nuclei arrangements and phonon spectrum near their minimum.

The pioneering Kramer's Transition State theory of chemical reactions \cite{hanggi1990reaction} describes the dynamics under thermal fluctuations between two local minima of the electronic ground state PES $V_0(\{R_n\})$ assumed to correspond to the two different chemical species before and after the chemical reaction. The kinetics of the chemical reaction is simply represented by the random path of a diffusive particle moving on the PES with friction and submitted to a random thermal force. The lowest saddle point of the PES (called in mathematics minimax) between the two minima corresponds to the transition state near which the most probable reaction paths should pass. It yields the energy barrier involved in the Arrhenius law.

The Frank-Condon principle, a well-known concept in photochemistry, explicitly incorporates the existence of PES. Subsequently, the absorption of a photon with a frequency of $\hbar \omega = V_j(\{R_n\})-V_i(\{R_n\})>0$ ($j>i$) initiates a direct transition from the initial PES $i$ to an upper PES $j$ at the same nuclear coordinates $\{R_n\}$. The next step is for the nuclei configuration to relax near the minimum of the new PES. Subsequently, the electronic state will also relax back to a lower PES, resulting in the emission of a photon with a frequency smaller than the initial value.

The BO approximation may break for a given PES in some domain of nuclei coordinates $\{R_n\}$ when the electronic gap between this PES and the nearest ones above and/or below. This is indicated by the fact that the energy differences $E_{i+1}(\{R_n\}) -E_{i}(\{R_n\})$ and/or $E_{i}(\{R_n\}) -E_{i-1}(\{R_n\})$ become sufficiently small to be within the range of phonon frequencies. The largest phonon frequency $\hbar \omega_c$ (frequency cut-off) in the phonon spectrum for realistic materials is usually about a fraction of $eV$. Then phonon quanta may have enough energy to trigger direct electronic transitions (at lowest order) between two PES. The consequence is that the electronic state of the system cannot stay invariant as assumed in the BO approximation.

Note that the electromagnetic spectrum does not exhibit any frequency cut-off (unlike the phonon spectrum).
Consequently, direct photonic transitions may be induced between different PES providing the photon frequency corresponds to the electronic gap. Since the external electromagnetic fields are weak compared to the microscopic fields, these transitions are usually described with the Fermi Golden rule.
Then the lifetime of the excited electronic state is sufficiently long at the scale of phonon frequencies so that the system has time to relax while staying on its PES (Frank-Condon principle).
The standard situations where the BO approximation holds concern insulators or semiconductors when the lowest electronic gaps are at least of the order of 1 $eV$ or much larger while the phonon excitations are at most fractions of $eV$.
Thus note that BO approximation is in principle not valid when the considered system is a gapless metal or nearly metallic.
Then, corrections to the BO approximation are generally taken into account by extra electron-phonon interactions. 
However keep in mind that in this paper we are considering transitions between localized electronic excitations in an infinite system. This assumption obviously requires that the global system is not metallic, but is a dielectric insulator. Then, the involved PES are still well described considering only a finite subsystem corresponding to nearby environment of the electronic excitation large enough but still finite. Nevertheless, assuming that the global system is infinite remains essential for explaining the energy disspation of the chemical reaction by phonon transportation. (However, note that in the case of periodic systems with electronic bands but with strong electron-phonon coupling, we may recover localized excitations such as polarons or excitons, the mobility of which could be described by a $N$-state model extending the 2-state model we study here but on a lattice).

Since $\mathcal{H}_e(\{R_n\})$ has infinitely many eigenenergies which become denser and denser at high energies, the spacing between two consecutive eigenenergies cannot always remain small for high energy electronic states so that BO approximation does not hold anymore for high energy PES. Of course, it is also not valid in the case of degenerate electronic states but this situation is rare because in a system with $N$ degrees of freedom, we generally have avoided crossing between the different PES due to level repulsion Fig. \ref{fig:pes_general_marcus} according to a Von Neumann-Wigner theorem also called avoided crossing theorem \cite{Neumann-Wigner} which is valid only for finite systems.

This theorem implies that the PES cannot be degenerate except at intersections between two PES but with dimension $N-2$ (instead of the expected dimension $N-1$) generally because of special symmetries. Such intersections are called conical, indeed in the case of electronic degeneracy, most distortions of the nuclei configuration raise linearly this degeneracy (except on a $N-2$ manifold).


\subsection{Extended Two-State BO approximation}
Kramers' theory may not be sufficient to describe the kinetics of all elementary reactions because it may involve a jump between two PES. The latter is a forbidden process within the BO approximation since it assumes that the system should remain on the same PES, see for example the lower scheme Fig. \ref{fig:pes_general_marcus} corresponds to the inverted regime in the Marcus theory.


Generally, it is more realistic to consider that the electronic subspace visited by the elementary chemical reaction path has dimension two since we expect the existence of two diabatic states - the initial and final ones, as previously explained. Fig. \ref{fig:pes_general_marcus} illustrates the approximate generation of two potential energy surfaces (PES) in the standard literature in chemistry. This is achieved by simply opening a gap at the intersection which are for example the two lowest possible energy surfaces $V_0(\{R_n\})$ and $V_1(\{R_n\})$. It should be noted that photochemically induced reactions may involve different PES.

This extended BO approximation becomes useful when the two PES come close enough to each other with a gap ranging in the phonon frequencies spectrum. It is equivalent to assume that the global wave function has the form
\begin{align}
    \Phi(\{R_n\},\{r_{\nu}\})= \phi_0(\{R_n\})\chi_0( \{R_n\}) + \phi_1(\{R_n\})\chi_1(\{R_n\})
    \label{genBO}
\end{align}
where $\chi_0(\{R_n\})$ and $\chi_1(\{R_n\})$ are the two consecutive eigenstates of the electronic Hamiltonian involved in the avoided PES crossing.

The matrix of operators $\hat{\mathbf{H}}$ is defined by Eqs. (\ref{Diag}) and (\ref{offdiag}) restricted in this 2d electronic subspace defined by $i=0$ and $i=1$ becomes a $2 \times 2$ matrix. The corrective term in Eq. (\ref{Diag}) is the same for both PES so that
\begin{align}
\begin{split}
    V_0(\{R_n\}) = E_0(\{R_n\}) + \sum_n \frac{\hbar^2}{2M_n} b_{0,1}^{(n) 2}(\{R_n\}) \\
    V_1(\{R_n\}) = E_1(\{R_n\}) + \sum_n \frac{\hbar^2}{2M_n} b_{0,1}^{(n) 2}(\{R_n\}) 
\end{split}\label{Diag2}
\end{align}

This $2\times 2$ matrix can be expanded based on Pauli matrices Eq. (\ref{pauli_matrix}) and the identity so that it gets the form
\begin{align}
\begin{split}
    &\hat{\mathbf{H}} = \mathcal{H}_K +A_0 (\{R_n\}) \\
    &\quad+A_x(\{R\quad_n\}) \sigma_x + A_y(\{R_n\},\{P_n\})) \sigma_y+ A_z(\{R_n\}) \sigma_z \\
\end{split}\label{spinbos}
\end{align}
where
\begin{align*}
    \begin{split}
    &A_0 (\{R_n\}) = \frac{1}{2} ( V_0(\{R_n\})+ V_1(\{R_n\}))\\
    &A_x (\{R_n\}) = C_{0,1}(\{R_n\}) = C_{1,0}(\{R_n\}) \\
    &\qquad\qquad\qquad\qquad\;\,= + \sum_n \frac{\hbar^2}{2M_n} b_{0,1}^{(n) 2} (\{R_n\}) \\
    &A_y (\{R_n\},\{P_n\}) = i D_{0,1}(\{R_n\},\{P_n\}) \\
    &\qquad\qquad\quad\;\,= i \sum_n \frac{\hbar }{2M_n} ( b_{0,1}^{(n)}(\{R_n\})P_n+P_n b_{0,1}^{(n)}(\{R_n\}) \\
    &A_z (\{R_n\}) =\frac{1}{2}( V_0(\{R_n\})-V_1(\{R_n\}))
    \end{split}
\end{align*}

Thus, the projection of the global Hamiltonian Eq. (\ref{gH}) in the subspace of two electronic eigenstates may be viewed as a Spin-Boson model where a fictitious spin $1/2$ describing the electronic state is coupled in all spin directions to the collection of nuclei. Note that the transverse fields $A_x(\{R_n\})$ and $A_y(\{R_n\})$ favor some hybridization of the electronic states associated with the two PES. Although this formulation is exact within the assumption of a two dimensional subspace of electronic wave functions, it is useful to redefine diabatic states because they closely represent the chemical structure.
It does not change formally the model to rotate the fictitious quantum spin $\mathbf{\sigma}$ that is to rotate the BO base of electronic state $\chi_0(\{R_n\}), \chi_1(\{R_n\})$. This rotation is accomplished by a unitary $2\times 2$ matrix, $\mathbf{U}(\{R_n\})$. The general form of this transformation may be expressed as the product of three rotations, each expressed in terms of the Pauli matrices defined in Eq. (\ref{pauli_matrix}).

\begin{align}
    \begin{split}
    &\mathbf{U}(\{R_n\}) = e^{i\gamma}(\cos \alpha+ i \sin \alpha\cdot \sigma_z) \cdot\\
    &\qquad\qquad\quad\:(\cos \theta + i \sigma_y \sin \theta )\cdot (\cos \beta+ i \sin \beta\cdot \sigma_z)
    \end{split}\label{unitmat}
\end{align}
where angles $\alpha,\beta,\gamma$ and $\theta$ are functions of $\{R_n\}$.

The standard schemes shown in Fig. \ref{fig:pes_general_marcus} suggest that when $\{R_n\}$ is near the first minimum of $V_0((\{R_n\})$ corresponding to the initial state, the electronic state $ \chi_D(\{R_n\})$ is nearly identical to $ \chi_0(\{R_n\})$ while $\chi_A(\{R_n\})$ is nearly identical to $\chi_1(\{R_n\})$.
Therefore, the unitary matrix should be the identity matrix, corresponding to $\alpha=\beta=\gamma=\theta=0$ in Eq. (\ref{unitmat}).
Similarly when $\{R_n\}$ is near the second minimum of $V_0(\{R_n\})$ corresponding to the final state where the electron is on the Acceptor site while $ \chi_A(\{R_n\})$, whereby $\chi_D(\{R_n\})$ should be nearly identical to $ \chi_1(\{R_n\})$ and $\chi_A(\{R_n\})$ should be approximately identical to $\chi_0(\{R_n\})$ so that the unitary matrix should be equal to $\sigma_x$ which corresponds to $\alpha= -\pi/4$, $\beta=\pi/4$, $\gamma=\pi/2$ and $\theta=-\pi/2$ in Eq.(\ref{unitmat}).
Thus, to generate diabatic states, the unitary transformation should vary from unity when $\{R_n\}$ is close to the first minimum of $V_0(\{R_n\})$, corresponding to a state with a well defined chemical formula to $\sigma_x$ when $\{R_n\}$ is close to the second minimum of $V_0(\{R_n\})$, corresponding to a state $V_0(\{R_n\})$ with another well defined chemical formula.

Such a transformation determines new states called diabatic as
\begin{align}
    \begin{pmatrix}\varphi_D \\ \varphi_A\end{pmatrix}=\mathbf{U}(\{R_n\})\cdot \begin{pmatrix} \varphi_0 \\ \varphi_1\end{pmatrix}
\end{align}

However since the conjugate operators $R_n$ and $P_n$ do not commute, the terms in $\hat{\mathcal{H}}$ are modified since the unitary matrix $\mathbf{U}$ depends on $\mathbf{U}(\{R_n\})$. Finally, the Hamiltonian retains the form of a spin-boson Hamiltonian with the form
\begin{align}
    \hat{\mathcal{H}} = \sum_{\alpha=0,x,y,z} \tilde{A}_{\alpha} (\{R_n\},\{P_n\}) \sigma_{\alpha}
    \label{Diabham}
\end{align}
where $\sigma_{\alpha}$ are the Pauli matrices defined Eq. (\ref{pauli_matrix}).
It is not necessary to explicitly show the complex result of the calculation of $ \tilde{A}_{\alpha} (\{R_n\},\{P_n\})$. 
The main issue is to find a unitary transformation $\mathbf{U}(\{R_n\})$ that is physically realistic. For this purpose, the new coefficients $\tilde{A}_{\alpha}(\{R_n\},\{P_n\})$ should have a smooth variation so that they can be approximated reasonably well by their lowest order expansions. It is not sure that this is possible in all cases for example when three Diabatic states are involved in the same elementary chemical reaction.
Then we should extend this analysis to a three-dimensional electronic subspace. The existence of a Diabatic representation is often assumed a priori in chemistry for example in the Marcus theory of electron transfer
and yields qualitatively correct results \cite{marcus2020electron}. This representation is very convenient for describing the electronic organization in terms of occupied or unoccupied electron orbitals and /or their quantum hybridization as done in chemistry.

\subsection{Semiclassical approximation}
Paul Dirac \cite{dirac1933lagrangian} proposed that classical mechanics could be derived from quantum mechanics as a consequence of destructive interference among paths in the classical phase space that do not externalize the Lagrangian action. 
Those with constructive interferences are merely near the classical trajectories. Dirac's ideas were later elaborated by R. Feynman in his path integral representation of quantum mechanics \cite{feynman1942}.
In contrast, the Heisenberg uncertainty principle asserts that the measurement accuracies for two conjugate variables, $R_n$ and $P_n$, must obey the rule $\Delta R_n, \Delta P_n \geq \hbar/2$. This rule gives the order of accuracy of the classical trajectory. Consequently, classical dynamics are only applicable to large displacements in the phase space, which are much larger than phonon quantum fluctuations.

In our system, electrons are very light and are fermions so they should be treated quantumly. However, the dynamics of the nuclei may be well studied within a semiclassical approximation subsequently to the Born-Oppenheimer approximation (\ref{BO_equation}) where the electrons are treated quantum mechanically and the nuclei classically.

However, the standard BO approximation may not be valid when two PES become too close to each other that is when some electronic frequencies $E_1(\{R_n\}) -E_0(\{R_n\}) \hbar \omega_e$ get into the phonon spectrum which extends up to some upper cut-off frequency $\omega_p$. We have shown above that such situations may be described by a Spin-Boson Hamiltonian (\ref{spinbos}) which takes into extra quantum interactions between these two PES and keeps a part of the electron dynamics.

\subsection{Conical Intersection of two potential energy surfaces}

The PES obtained with the BO approximation are generally difficult to calculate and moreover uneasy to use for understanding the corresponding nuclei configurations as well as the possible chemical reactions \cite{farfan2020systematic, schlegel2003exploring}. The semiclassical approximation is similar to a gauge transformation

\begin{align}
    \begin{pmatrix}\chi_D(\{R_n\})\\ \chi_A(\{R_n\})\end{pmatrix}=\begin{pmatrix}\cos \alpha(\{R_n\}) & \sin \alpha(\{R_n\}) \\ -\sin \alpha(\{R_n\})& \cos \alpha(\{R_n\})\end{pmatrix}\cdot\begin{pmatrix}\chi_0(\{R_n\})\\ \chi_1(\{R_n\})\end{pmatrix}
\end{align}

To obtain a good diabatic base, the new coefficients of the Spin-Boson system  Eq. (\ref{spinbos}) should
be a function of the nuclei coordinate which is as smooth as possible and well approximated by low order expansion since this is not true. However, there are no standard criteria for optimizing the choice of this rotation angle $\alpha(\{R_n\})$ for obtaining the \textit{best} Diabatic base.

We assume that there exist two orthogonal electronic states so that $\chi_D(\{R_n\})$ should well represent the initial state of the system where the electron only occupies the orbital on the Donor (which is deformable when $\{R_n\}$ vary) while $\chi_A(\{R_n\})$ represents the final state of the system where the electron only occupies the orbital on the Acceptor.

\begin{align}
    \hat{\mathcal{H}} = \mathcal{H}_K +A_0 (\{R_n\})+ A_x^{\prime} (\{R_n\})  \sigma_x + A_z^{\prime}(\{R_n\}) \sigma_z \label{spinbosD}
\end{align}
where
\begin{align}
    \begin{pmatrix} A_x^{\prime} (\{R_n\}) \\ A_z^{\prime} (\{R_n\})\end{pmatrix}=\begin{pmatrix}\cos{\alpha(\{R_n\})} & \sin{\alpha(\{R_n\})} \\ -\sin \alpha(\{R_n\})& \cos \alpha(\{R_n\}\end{pmatrix}\cdot \begin{pmatrix}A_x (\{R_n\}) \\ A_z (\{R_n\})\end{pmatrix}
\end{align}

Assuming we got a good Diabatic representation, we see that when neglecting $ A_x^{\prime}(\{R_n\})$, coefficients $ A_z^{\prime}(\{R_n\})$ favors a spin $\vert{\uparrow}\rangle$ or $\vert{\downarrow}\rangle$ corresponding to the electronic
state $\chi_D(\{R_n\})$ or $\chi_A(\{R_n\})$. We get the purely ionic case.
Conversely when neglecting $A_z^{\prime}(\{R_n\})$, coefficients $ A_x^{\prime}(\{R_n\})$ favors a purely covalent state$1/2( \vert{\uparrow}\rangle+\vert{\downarrow}\rangle)$ or $1/2( \vert{\uparrow}\rangle+\vert{\downarrow}\rangle)$

More precisely potential $A_0 (\{R_n\}$ is supposed to have a single minimum which may be assumed to be the origin for the nuclear coordinates as well as for the energy scale. Then it is assumed to be well approximated by its expansion as a quadratic function
\begin{align}
    A_0 (\{R_n\} \approx \frac{1}{2} \sum_{n,m} a_{n,m} R_n X_m
    \label{ham0}
\end{align}
Next, we can expand to the lowest order
\begin{align}
    A_x^{\prime} (\{R_n\}) \approx  A_x^{\prime} (\{0\})+ \sum_n t_n R_n \label{hamx}\\
    A_z^{\prime} (\{R_n\}) \approx  A_z^{\prime} (\{0\})+ \sum_n h_n R_n \label{hamz}
\end{align}

When the electron remains on the Donor orbital which corresponds to the spin $\vert{\uparrow}\rangle$, we get a PES we call Diabatic PES $\vert{\uparrow}\rangle$ the Diabatic PES is $V_D(\{R_n\})= A_0 (\{R_n\}) + A_z^{\prime}(\{R_n\}) $ while when the electron remains on the acceptor orbital represented by $\vert{\downarrow}\rangle$, the corresponding Diabatic PES
is $V_A(\{R_n\})= A_0 (\{R_n\}) - A_z^{\prime}(\{R_n\})$. The description used in \cite{aubry2014semiclassical} assumes the existence of this base of orbitals as already done in the Marcus theory of Electron Transfer.

\section{ A simple prototype model with Ultrafast Targeted Electron Transfer}

The phenomenon of exceptional chemical reactions that do not obey the Arrhenius law is ubiquitous in biological systems and plays an essential role in the functioning of life. A lot of work has been done to describe fast electron transfer in the different systems \cite{kopidakis2001targeted, aubry2001analytic, maniadis2004classical, maniadis2005targeted, memboeuf2005targeted}. For example, UET at the Photosynthetic Reaction Center (PRC) allows living photosynthetic cells to capture sunlight energy with great efficiency, which is then converted into chemical energy for subsequent use in biological processes \cite{kopidakis2001targeted}.
We believe that the novel approach will prove beneficial in elucidating a multitude of puzzling phenomena observed in living cells, thereby stimulating further research to advance these new concepts.

The extension of electron transfer occurs in the vicinity of the situation where the PES exhibits a degenerate ground state, which continuously connects the state with the electron on the Donor state. This phenomenon is influenced by the competition between ionic and covalent interactions.

Our approach for electron transfer consists of studying the quantum dynamics of the wave function of an electron (or any other kind of quantum excitation) from a donor site $D$ to an acceptor site $A$.
The wave function of the electron at time $t$ has the form $\varphi_D(t)\vert{D}\rangle + \varphi_A(t)\vert{A}\rangle$ where $\vert{D}\rangle$ is the orbital of the electron localized at the Donor site $D$ and $\vert{A}\rangle$ the orbital of
the electron localized at the Acceptor site $A$. The Hamiltonian of such a model has the simple form
\begin{align}
    H = E_D |\varphi_D|^2+E_A |\varphi_A|^2 +\Gamma
    (\varphi_D^{\star} \varphi_A + \varphi_A^{\star}\varphi_D )
    \label{dimer}
\end{align}
where $E_D$ and $E_A$ are the onsite energies on the Donor and Acceptor and $\Gamma$ is the transfer integral which depends on the overlap between the two orbitals on the Donor and Acceptor. However, this model is not isolated. It should be taken into account for its interaction with the complex environment in which the electron transfer couples.

Our purpose is to produce an example of model where TET could occur, it is convenient to assume for simpliciity that the model we consider is nearly symmetric which implies the reaction energy $E_D-E_A$ is relatively small as it is in elementary biochemical reactions. ATP hydrolysis reaction energy 0.3 eV may be considered as a good unit for scaling much smaller than those of most chemical reactions in inorganic chemistry.

More precisely, the onsite energies $E_D$ and $E_A$ depend on this environment because a charge transfer between Donor and Acceptor changes the local electric field which polarizes the environment and consequently generates nuclei displacement. Otherwise, since organic molecules are easily deformable, the spatial distance
between the donor and acceptor may also vary during electron transfer, which changes the orbital overlap and consequently the transfer integral $\Gamma$.

It can be assumed that the environment is described as a collection of infinitely many coupled harmonic oscillators. Two classical linear oscillators are selected, with variables $u_z$ and $u_x$, which are coupled to the other oscillators. These oscillators will be considered as a Langevin bath.

The mode $u_z$ describes the deformation of the environment essentially due to charge transfer.
We assume that it is linearly coupled to both charge densities $|\varphi_D|^2$ and $|\varphi_A|^2$ but since $|\varphi_D|^2+|\varphi_A|^2=1$ is constant, it turns out to be only coupled to $C=|\varphi_A|^2-|\varphi_D|^2$. Thus, the variation of $u_z$ changes linearly the effective electronic levels $E_D$ and $E_A$.
Exchanging Donor and Acceptor, $C$ is changed into $-C$, hence it may be argued that this mode is antisymmetric in relation to the electron exchange between Donor and Acceptor. 
Mode $u_x$ represents the deformation of the environment when the spatial distance between the Donor and the Acceptor sites varies which thus changes the overlap integral $\Gamma$.
We assume again this oscillator is linearly coupled to $ (\varphi_D^{\star} \varphi_A + \varphi_A^{\star}\varphi_D )$. Unlike mode $u_z$, this mode $u_x$ is symmetric when exchanging Donor and Acceptor and thus must be different from the symmetric mode $u_z$.
The Hamiltonian of our model (when mode $u_z$ and $u_x$ are decoupled from the Langevin bath)

\begin{align}
    \begin{split}
        H=E_D|\varphi_D|^2 &+ E_A |\varphi_D|^2+ k_zu_z(|\varphi_A|^2-|\varphi_D|^2)\\
        & + (\epsilon_x+k_xu_x)(\varphi_D^\ast\varphi_A+\varphi_A^\ast\varphi_D )   \\
        & +\frac{1}{2}p_z^2 +\frac{1}{2}\Omega_z^2u_z^2+\frac{1}{2}p_x^2+\frac{1}{2}\Omega_x^2u_x^2
        \label{hamiltonian}
    \end{split}
\end{align}
and its Hamilton equations are
\begin{align}
    \begin{split}
        & i\hbar\dot{\varphi}_{D} =(E_{D}-k_z u_z)\varphi_{D}+(\epsilon_x+k_x u_x)\varphi_{A} \\
        & i\hbar\dot{\varphi}_{A}  =(E_{A}+k_z u_z)\varphi_{A}+(\epsilon_x+k_x u_x)\varphi_{D} \\
        & \ddot{u}_z+\Omega^{2}_{z}u_z +k_z(\left\lvert \varphi_{A}\right\rvert^2 - \left\lvert \varphi_{D}\right\rvert^2) = 0  \\
        & \ddot{u}_x++\Omega^{2}_{x}u_x +k_x(\varphi^{*}_{D}\varphi_A+\varphi^{*}_{A}\varphi_D) = 0
        \label{equation_without_noise}
    \end{split}
\end{align}

The quantity $\hbar$ will be dropped from these equations. Consequently, $1/\hbar$ will be chosen as the unit of time, with $\hbar$ as the unit of energy, and the variables $u_z$, $u_x$, as well as the energies $E_D$, $E_A$, and frequencies $\Omega_z$, $\Omega_x$ will be rescaled.
If we now presume that phonon modes $u_x$ and $u_z$ are coupled to two independent Langevin baths (symmetric and antisymmetric) which add damping and Langevin noise to the oscillator equation Hamilton equations.

\begin{align}
    \begin{split}
        & i\dot{\varphi}_{D} =(E_{D}-k_z u_z)\varphi_{D}+(\epsilon_x+k_x u_x)\varphi_{A}                                                               \\
        & i\dot{\varphi}_{A}  =(E_{A}+k_z u_z)\varphi_{A}+(\epsilon_x+k_x u_x)\varphi_{D}                                                             \\
        & \ddot{u}_z+\gamma_z\dot{u}_z+\Omega^{2}_{z}u_z +k_z(\left\lvert \varphi_{A}\right\rvert^2 - \left\lvert \varphi_{D}\right\rvert^2) = \eta_{z}(t)  \\
        & \ddot{u}_x+\gamma_x\dot{u}_x+\Omega^{2}_{x}u_x +k_x(\varphi^{*}_{D}\varphi_A+\varphi^{*}_{A}\varphi_D) = \eta_{x}(t)
        \label{mail_set_of_equations}
    \end{split}
\end{align}

Since we assumed above a quasi symmetry for the Donor Acceptor system, we should also split the surrounding phonon collection
into symmetric or antisymmetric modes thus generating two almost non interacting Langevin baths, the antisymmetric one interacting only with the antisymmetric ionic mode $u_z$ and the symmetric one with the covalent mode $u_x$. $\gamma_z$ is a constant that depends on the coupling of this oscillator $z$ which is the Langevin bath and the same for the $x$ component. $\eta_x(t+\tau)$ and $\eta_z(t)$ are random Gaussian white noise at temperature $T$ with the standard correlation:
\begin{align*}
     & \langle \eta_x(t+\tau)\eta_x(t)\rangle = 2\gamma_x k_B T \delta(\tau) \\
     & \langle \eta_z(t+\tau)\eta_z(t)\rangle = 2\gamma_z k_B T \delta(\tau)
\end{align*}
\begin{figure*}[t]
    \begin{tabular}{ccc}
        \includegraphics[width=0.6\columnwidth]{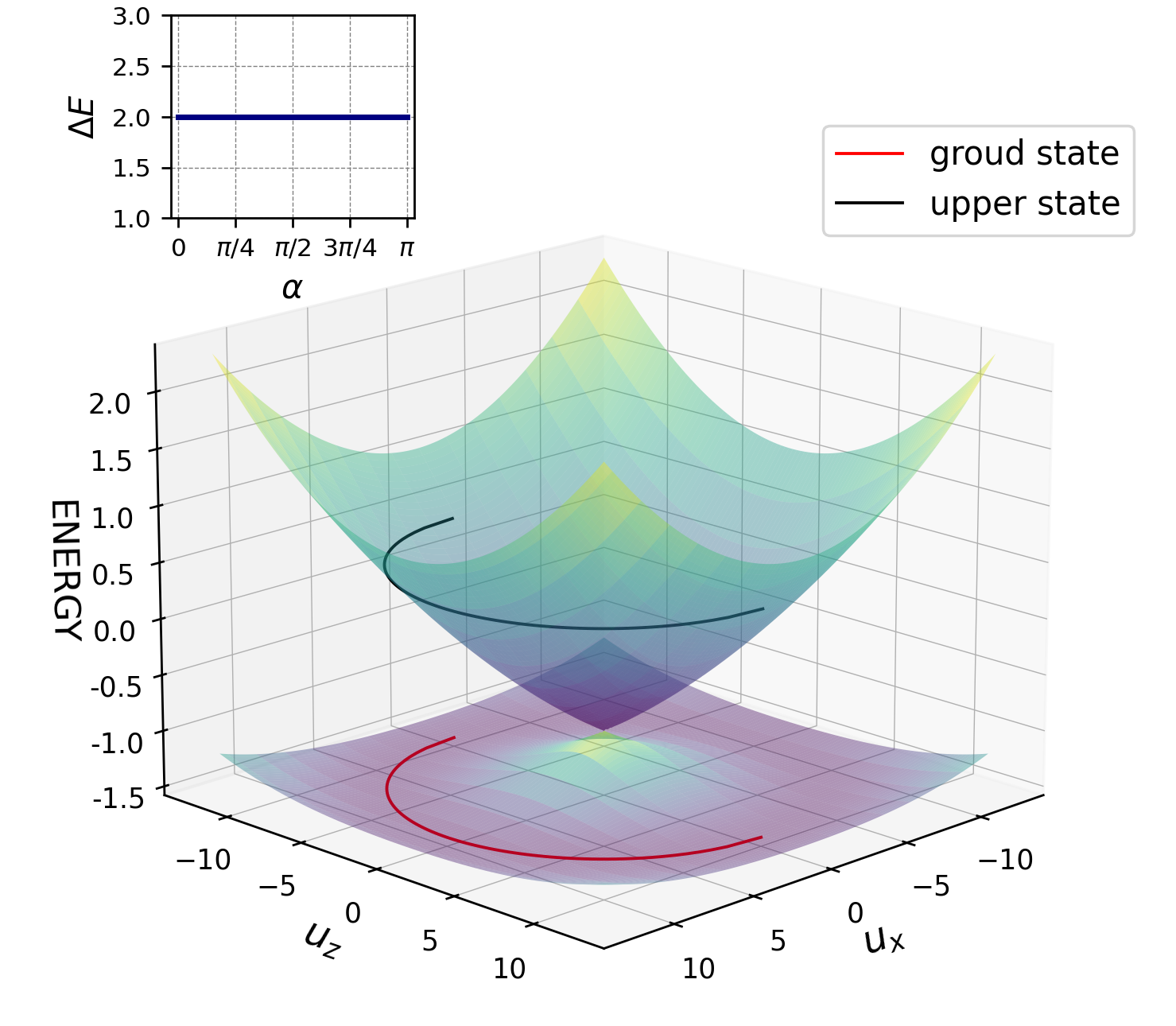}  &
        \includegraphics[width=0.6\columnwidth]{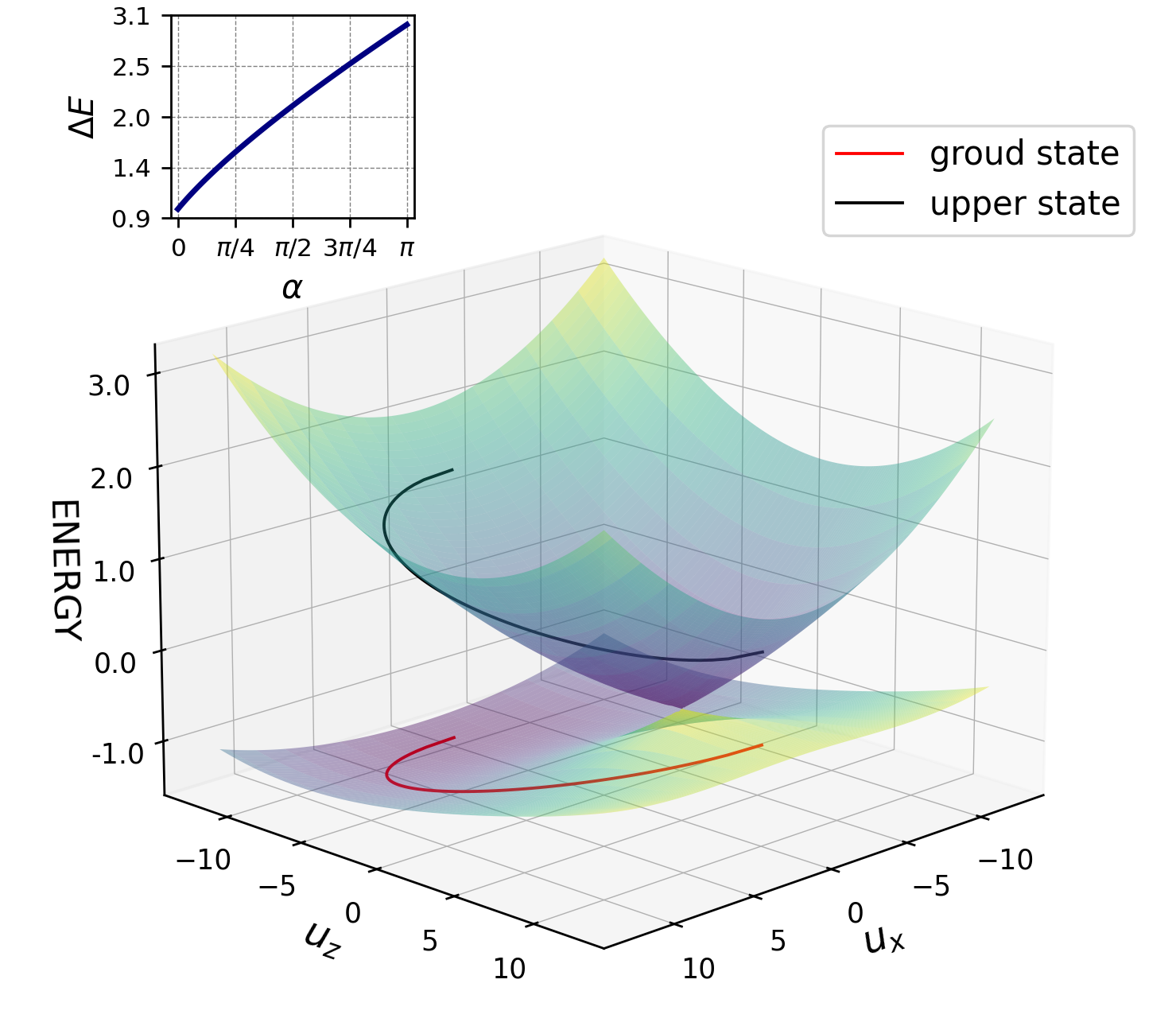} &
        \includegraphics[width=0.6\columnwidth]{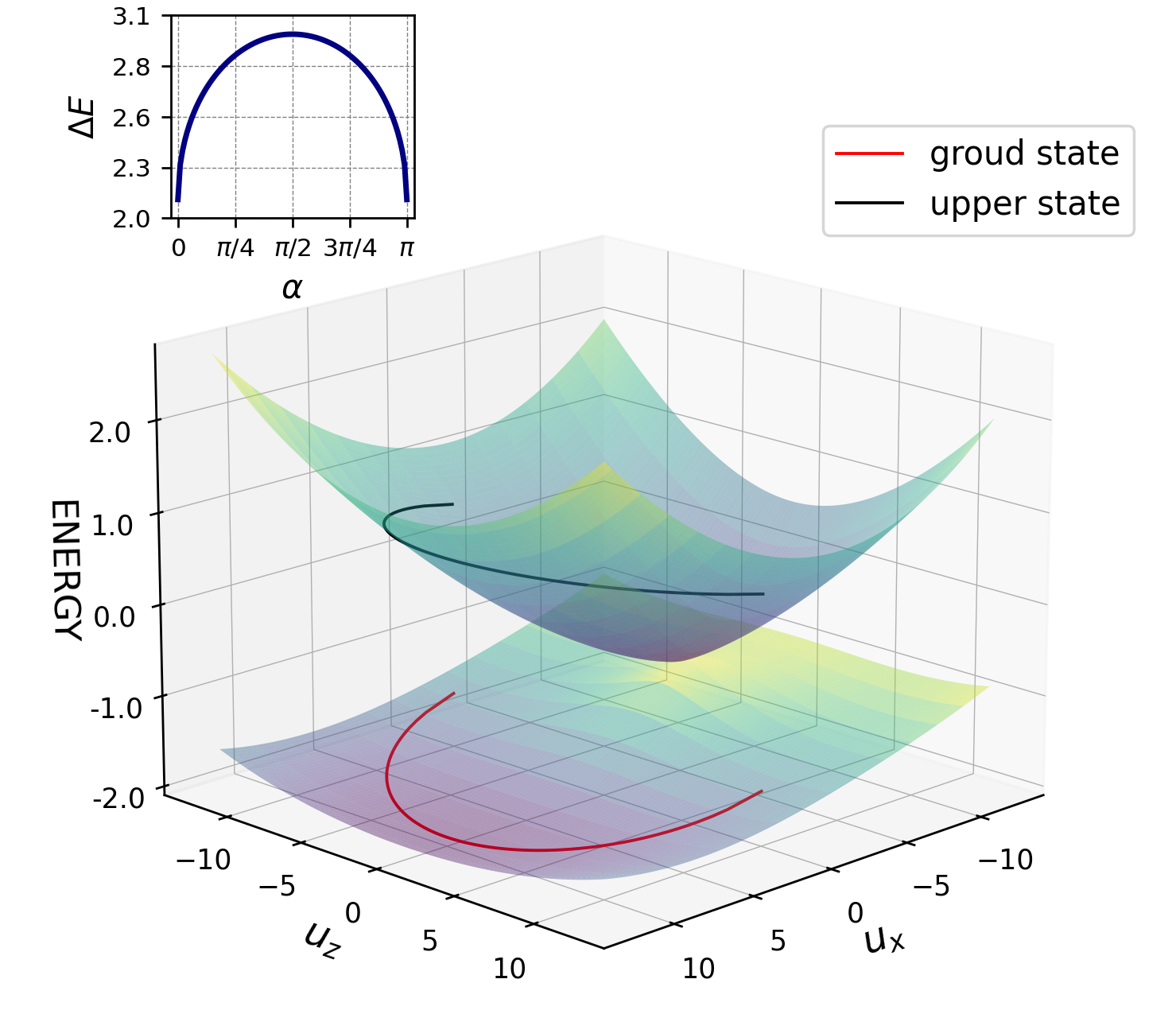}  \\
        
        \includegraphics[width=0.6\columnwidth]{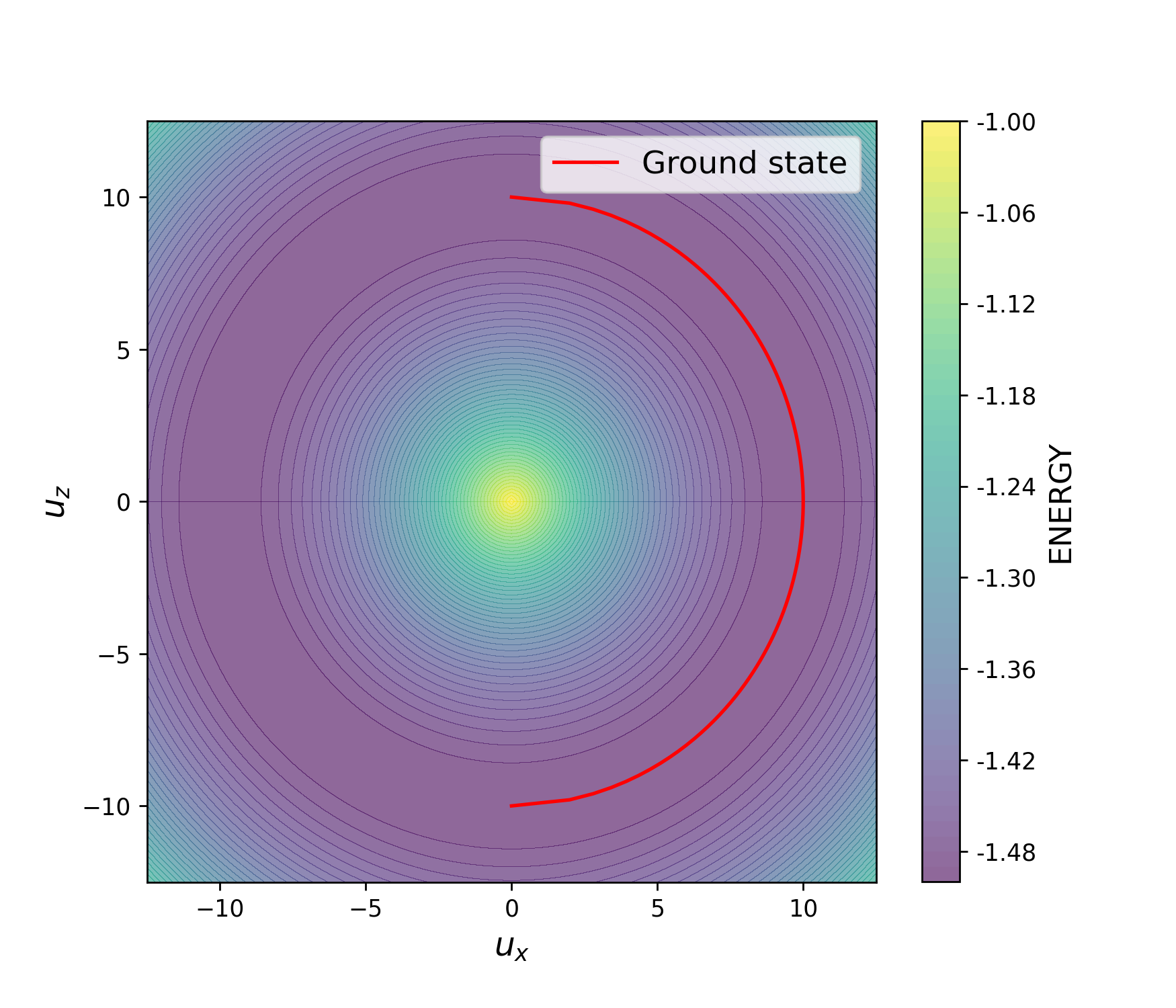}  &
        \includegraphics[width=0.6\columnwidth]{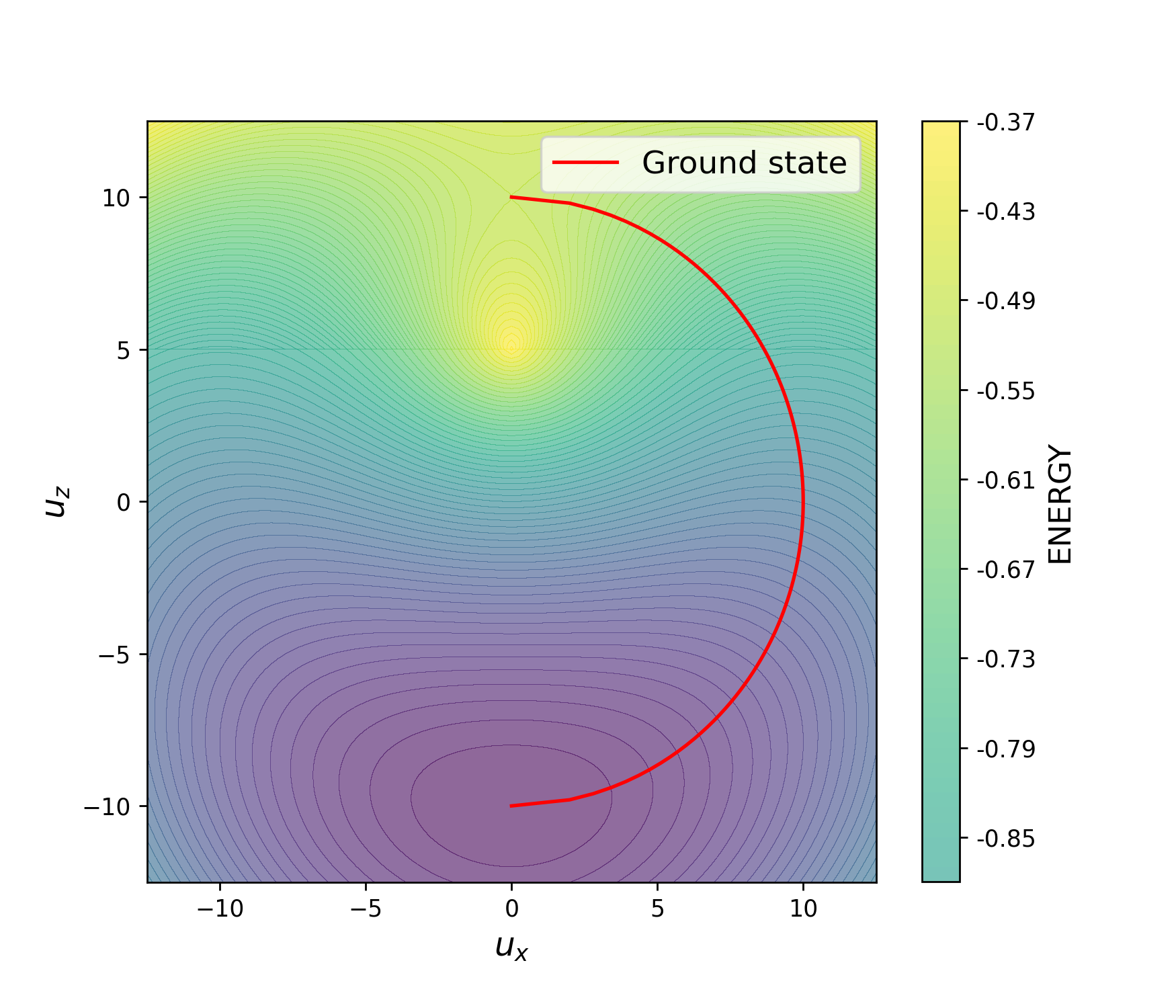} &
        \includegraphics[width=0.6\columnwidth]{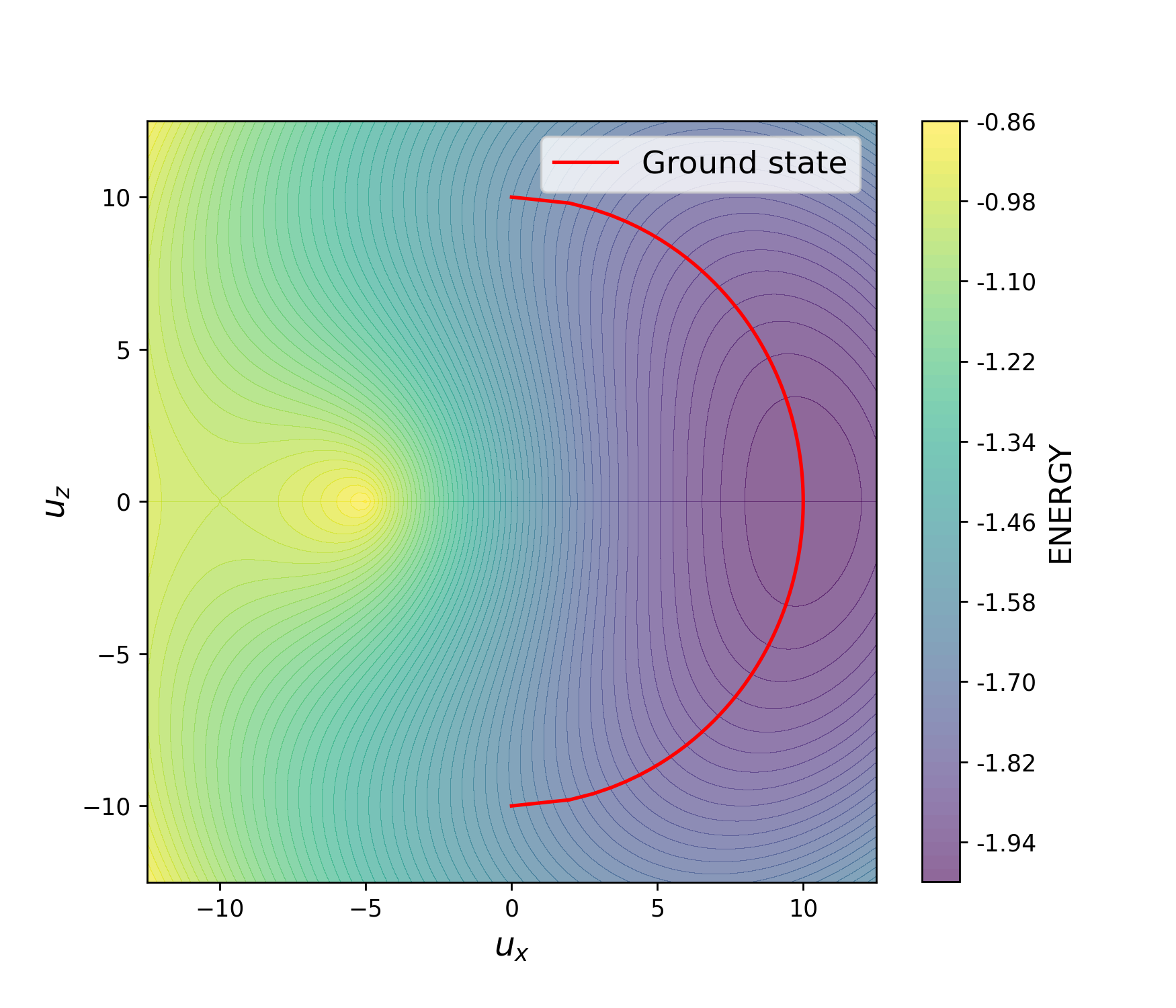}  \\
    \end{tabular}
    \centering
    \caption{Potential energy surfaces over modes $u_x$ and $u_z$.
        The red and black curves illustrate the reaction path of
        the lower and upper levels, respectively.
        The first left figure represents pure TET conditions
        ($E_A=E_D$, $\epsilon_x=0$ and
        $k_x^2/\omega_x^2 =k_z^2/\omega_z^2 \to 1=1$).
        The middle figure shows the shift in
        the conical intersection between two PESs 
        ($E_A\neq E_D$, $| E_A-E_D|>0$, $\epsilon_x = 0$ and
        $k_x^2/\omega_x^2 =k_z^2\omega_z^2\to 1=1$).
        The right figure shows the displaces of intersection over the $u_x$ mode near TET conditions
        ($E_A= E_D$, $\epsilon_x\neq 0$ and
        $k_x^2/\omega_x^2 =k_z^2\omega_z^2\to 1=1$).}
    \label{fig:table_TET_PES}
\end{figure*}

This model is equivalent to a Spin-Boson model which corresponds to a quantum spin $1/2$ with quantum state $\varphi_D \vert{\uparrow}\rangle + \varphi_A \vert{\downarrow}\rangle$ coupled to a phonon bath through variables $u_z$ and $u_x$. If we vanish the coupling $k_x$ and keep $\epsilon_x$ small, our model becomes equivalent to the standard Marcus model which is well-known in chemistry for describing electron transfer. We then recover the Arrhenius law. The covalent coupling described by parameter $k_x$ is usually dropped.
This model from the first principles was described in detail \cite{aubry2014semiclassical}.

At zero degree where $\eta_z(t)=\eta_x(t) =0$, equations (\ref{mail_set_of_equations}) readily yield
\begin{align}
    \dot{H} = -\gamma_z \dot{u}_z^2 -\gamma_x \dot{u}_x^2
    \label{dissip}
\end{align}
proving that the system energy decays as a function of time because the oscillators are damped.
By substitution in Eqs. (\ref{equation_without_noise}), we obtain a DNLS dimer Hamiltonian $H_{aa}$ which may be viewed as the anti adiabatic approximation of the general Hamiltonian valid when suppose the atoms are very light that is when the frequencies $\Omega_z$ and $\Omega_x$ are very large (so that the nuclei follow adiabatically the electronic variables).
\begin{align}
\begin{split}
    H_{aa} = &- \frac{1}{2} \frac{k_z^2}{\Omega_z^2}(|\varphi_A|^2 -|\varphi_D|^2)^2 -\frac{1}{2}\frac{k_x^2}{\Omega_x^2}(\varphi_D^{\star} \varphi_A + \varphi_A^{\star}\varphi_D )^2 \\
    &+E_D |\varphi_D|^2+E_A |\varphi_A|^2+\epsilon_x (\varphi_D^{\star} \varphi_A + \varphi_A^{\star}\varphi_D ) \label{aaham} 
\end{split}
\end{align}
    
Since $|\varphi_D|^2 +|\varphi_A|^2=1$ is a time-invariant of the first two Eqs.(\ref{equation_without_noise}), we can redefine variable $\rho= | \varphi_A|^2$ which only varies between $0$ and $1$ (charge transfer)
so that $\varphi_A= \sqrt{\rho} e^{i\alpha_A}$ and $\varphi_D= \sqrt{1-\rho} e^{i\alpha_D}$ where $\alpha_A$ and $\alpha_D$ are phase variables. Then the equilibrium state are minimum of
\begin{align}
    \begin{split}
        H_{aa}  &= - \frac{1}{2} \frac{k_z^2}{\Omega_z^2} (2\rho -1)^2 - \frac{2k_x^2}{\Omega_x^2} \rho(1-\rho) \cos^2 (\alpha_A-\alpha_D) \\
    &+ E_D (1-\rho) +E_A \rho +2 \epsilon_x \sqrt{\rho(1-\rho)} \cos (\alpha_A-\alpha_D)
    \end{split}
\end{align}

Minimizing $H_{aa}$ with respect to the phase yields $\alpha_A-\alpha_D =0 \mod 2\pi$ when $\epsilon_x<0$ and $\alpha_A-\alpha_D = \pi \mod 2\pi$ so that the minimum of $H_{aa}$ is obtained by minimizing
\begin{align}
    \begin{split}
        F(\rho) &= - \frac{1}{2}\frac{k_z^2}{\Omega_z^2} (2\rho -1)^2 -\frac{2k_x^2}{\Omega_x^2} \rho(1-\rho) + E_D (1-\rho) \\
        &\qquad\qquad\qquad\quad\:+E_A \rho - 2 |\epsilon_x| \sqrt{\rho(1-\rho)}\\
        &= \left(  \frac{2 k_x^2}{\Omega_x^2} - \frac{2 k_z^2}{\Omega_z^2}\right)(\rho^2 -\rho)+(E_A-E_D) \rho \\
        &\qquad\qquad\qquad\quad\:- 2 |\epsilon_x| \sqrt{\rho(1-\rho)} + E_D- \frac{1}{2} \frac{k_z^2}{\Omega_z^2} 
    \end{split}
\end{align}

The ground state of $H_{aa}$ is degenerate when
\begin{align}
    \begin{split}
        &E_A=E_D\\
        &\frac{k_z^2}{\omega_z^2} = \frac{k_x^2}{\omega_x^2}\\
        &\epsilon_x=0
        \label{tet_conditions}
    \end{split}
\end{align}

TET should be searched in the vicinity of this set of parameters Eq. \ref{tet_conditions}.

This model exhibits a physical flaw due to the fact that the transfer integral $\Gamma = \epsilon_x e^{k_x u_x/\epsilon_x}$ varies exponentially as a function of the spatial distance $u_x$ between the two orbitals. 
The linear expansion $\Gamma(u_x) = \epsilon_x + k_x u_x$ is physically acceptable on condition that it does not change sign when $u_x$ varies.
Otherwise, the resulting artifacts may be unacceptable.
In order to obtain realistic scenarios, it is necessary to verify that the transfer integral does not undergo a change of sign during the time evolution of the system. Such a situation arises in the overdamped regime when the damping coefficients, $\gamma_z$, and $\gamma_x$ are sufficiently large. Alternatively, the exponential form may be employed. 

The first problem is to detect irreversible TET at zero temperature around this set of parameters and to explore its domain of existence. We expect that the fastest TET is obtained in the crossover region between the underdamped region where TET oscillates a long time between Donor and Acceptor and the overdamped region where the large damping slows down TET. The second problem is to investigate the effect of temperature. When TET does not occur at zero temperature, it might take place at an optimized temperature which obeys the Markus theory and the Arrhenius law can be observed.


\section{Results}

\subsection{Potential Energy Surface}

First, it is crucial to comprehend the landscape of the reaction by calculating the potential energy surfaces. By employing this methodology, it is possible to ascertain the molecular dynamics, chemical reaction pathways, and other chemistry, physics, and biology-related processes and phenomena at the atomic scale through the use of an accurate PES. The PES describes the variation of the energy of a molecule as a function of the nuclear coordinates.

The accurate PES can be derived by minimizing Hamiltonian Eq. (\ref{aaham}). The solution of the dynamical equations (\ref{mail_set_of_equations}) always converges to a stationary point of the Hamiltonian for any initial condition, which represents time-independent solutions of Eqs. (\ref{equation_without_noise}). As a result, the initial values for both modes are

\begin{align*}
     & u_x=-\frac{k_x}{\Omega^{2}_x}(\varphi^{\star}_D \varphi_A+\varphi^{\star}_A \varphi_D) =0.0  \\
     & u_z=-\frac{k_z}{\Omega^{2}_z}(\left\lvert \varphi_A\right\rvert^2 -\left\lvert \varphi_D\right\rvert^2 )
\end{align*}\label{init_ux_uz}

If we consider the dimer problem, there are two scenarios to examine: one involving electron transfer and the other aligning with the traditional framework of Marcus' theory for both normal and inverted cases. Achieving ultrafast electron transfer requires careful satisfaction of TET conditions as defined in Eqs. (\ref{tet_conditions}) along with initial parameters meeting equation (\ref{init_ux_uz}). Fig. \ref{fig:table_TET_PES} demonstrates the PES with conical intersection under various TET regime parameters and projects the ground state onto modes $u_x$ and $u_z$. The PES is shown in  Fig. \ref{fig:table_TET_PES}, where the red and black lines represent the reaction pathway in the ground and excited states, respectively. Initial system parameters ($k_x$, $k_z$, $\omega_x$, $\omega_z$, $\gamma_x$, $\gamma_z$) remain consistent across each PES, while only $E_A$, $E_D$, $\epsilon_x$ vary. In this instance, a temperature equal to 0 K (without random force) was utilized for these results obtained through Python calculations.

\begin{figure}[ht]
\centering
\setlength{\lineskip}{-4mm}
    \includegraphics[width=0.9\columnwidth]{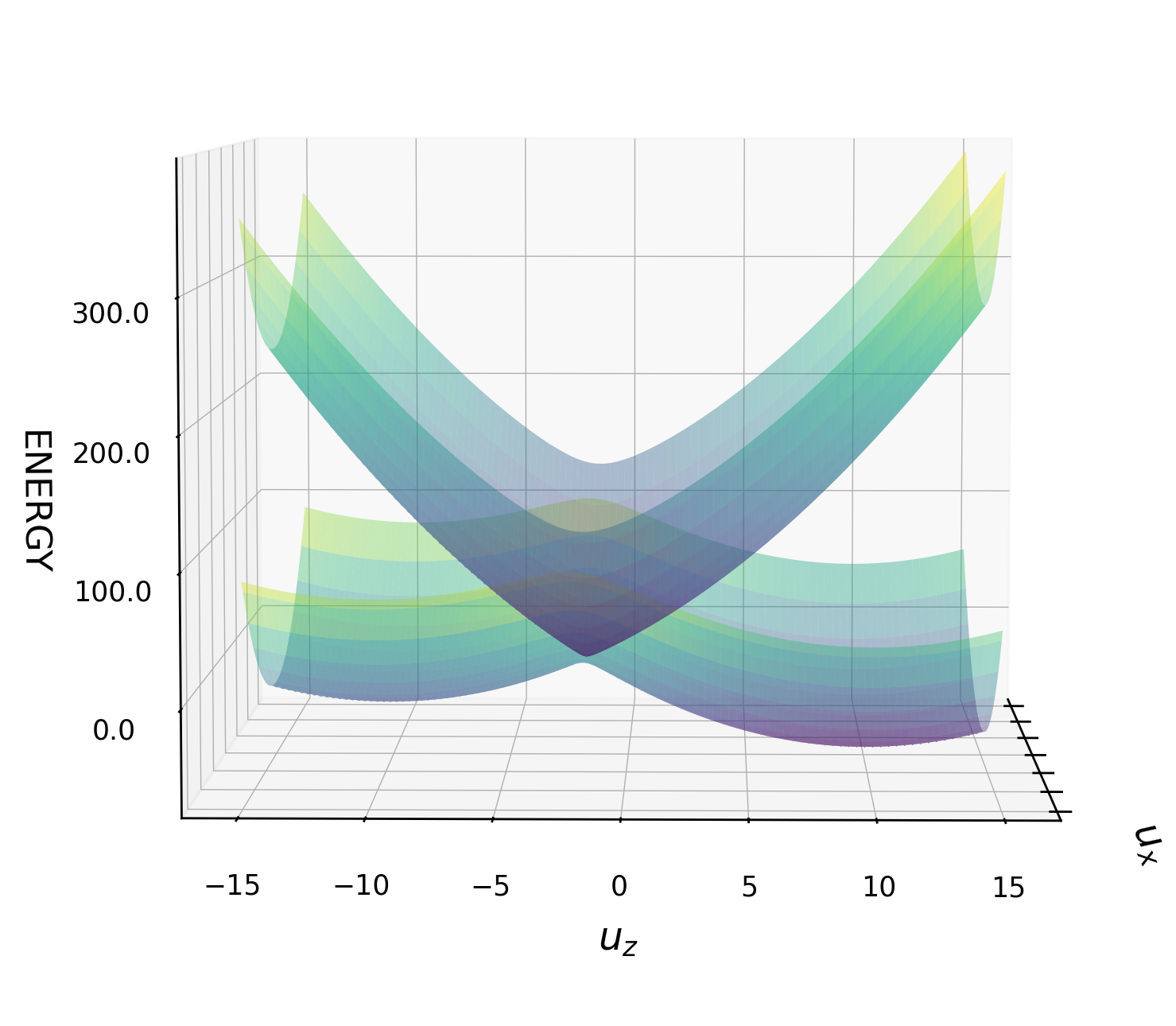} \\%
    \includegraphics[width=0.9\columnwidth]{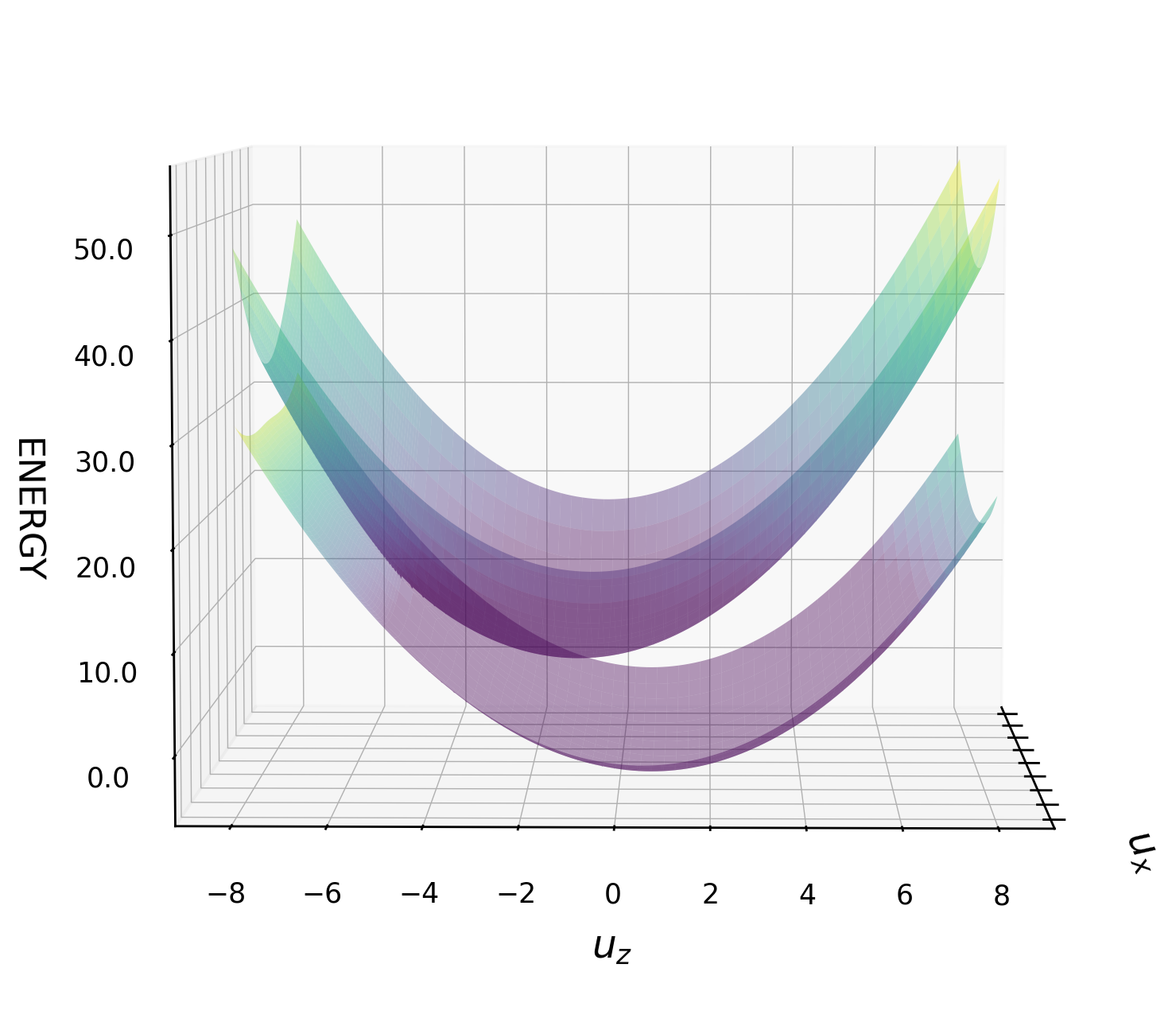}
    \caption{The PES of the normal and inverted case of Marcus' theory over the modes $u_x$ and $u_z$}%
    \label{fig:away_tet_pes}
\end{figure}

\begin{figure*}[ht]
    \begin{tabular}{cc}
        \includegraphics[width=0.7\columnwidth]{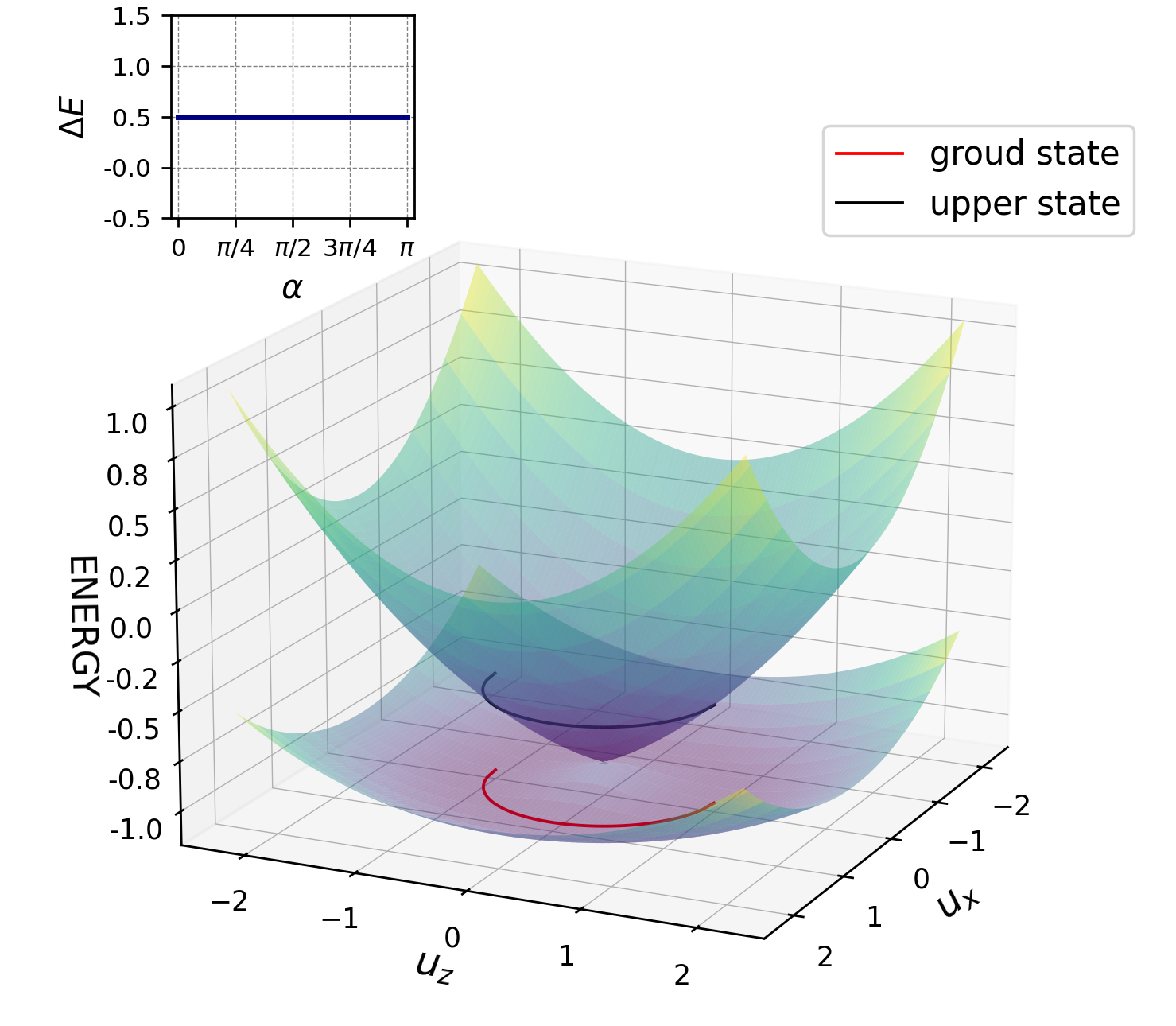}  &
        \includegraphics[width=1.1\columnwidth]{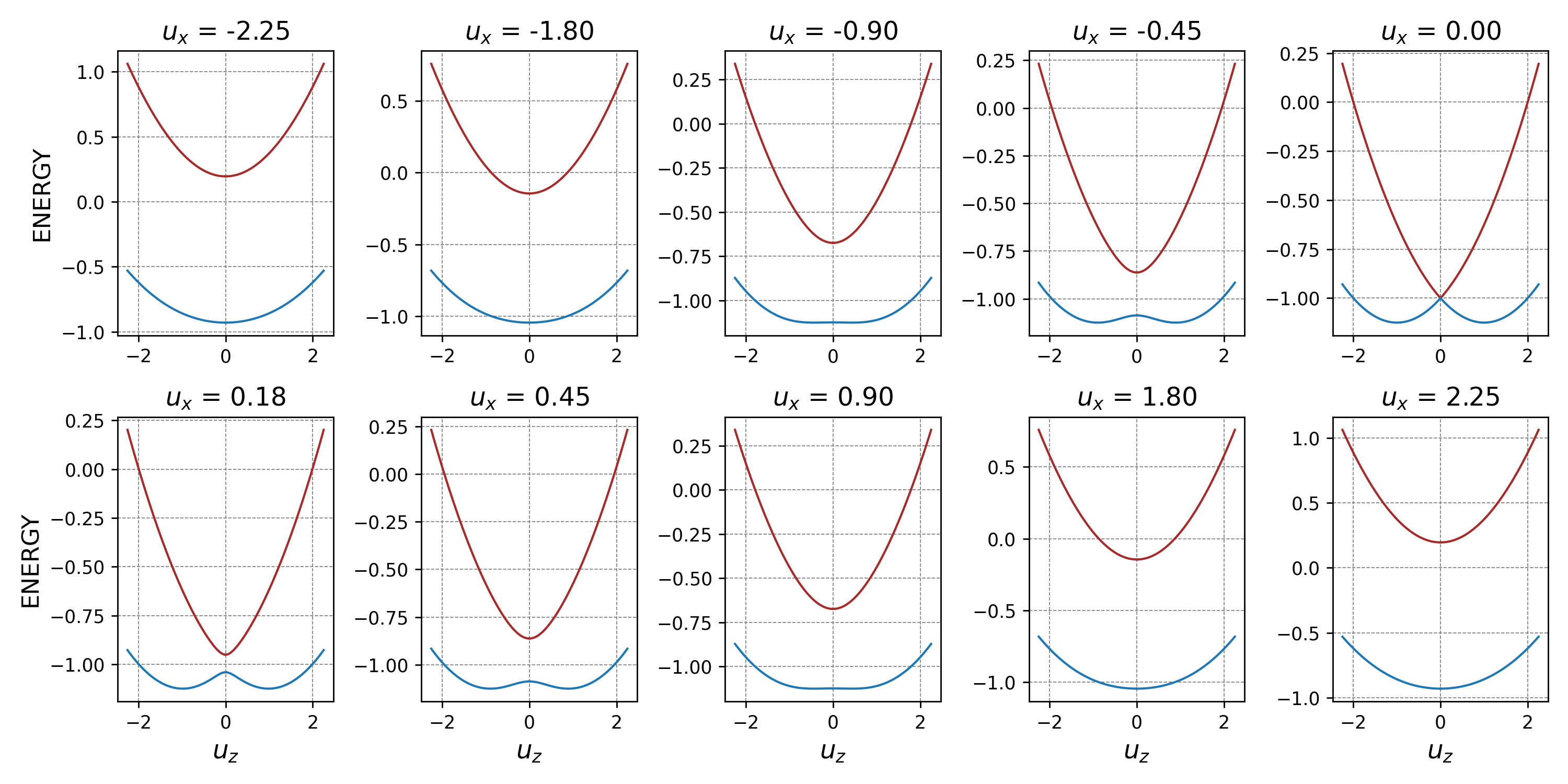} \\
        \includegraphics[width=0.7\columnwidth]{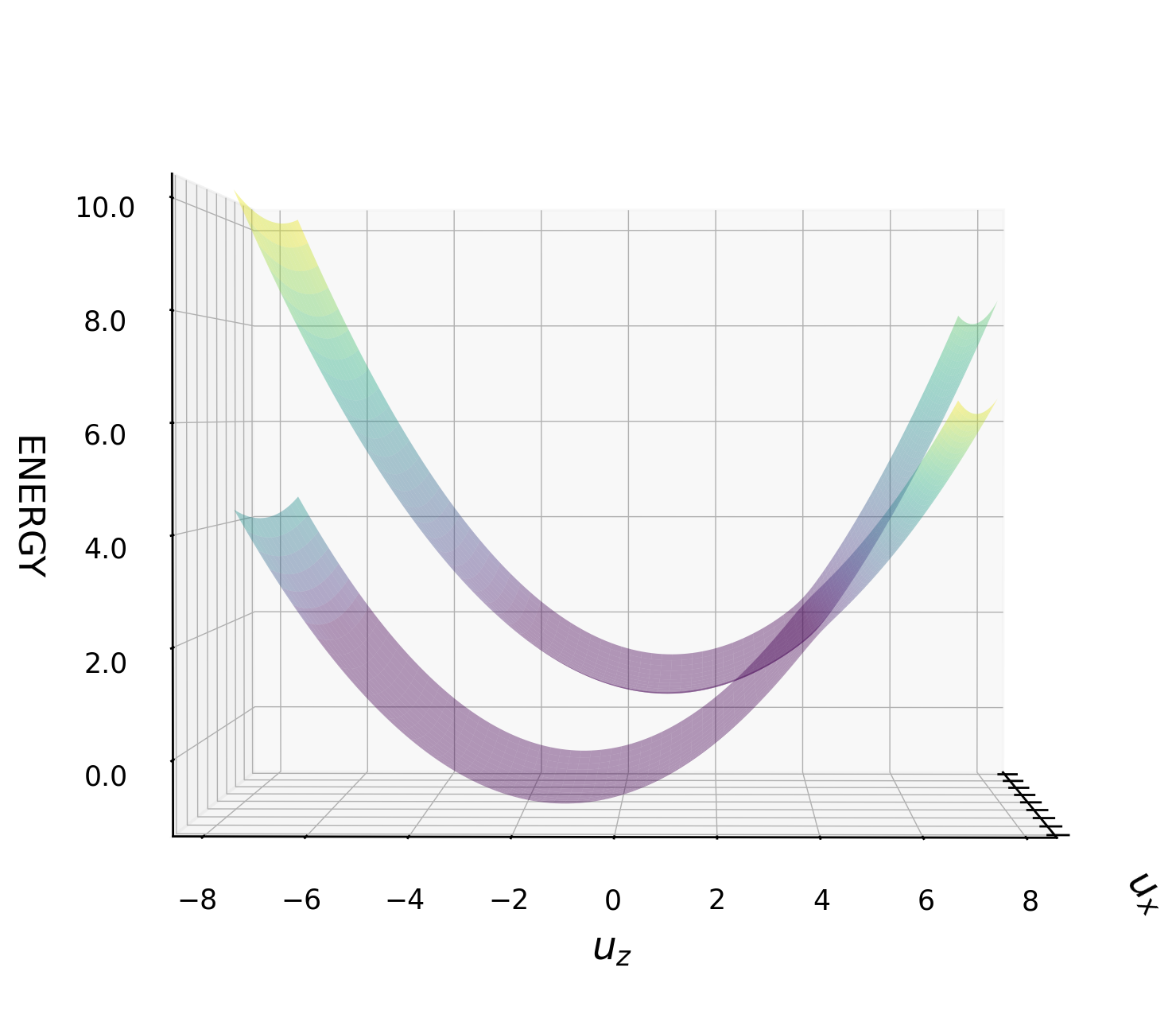}  &

        \includegraphics[width=1.1\columnwidth]{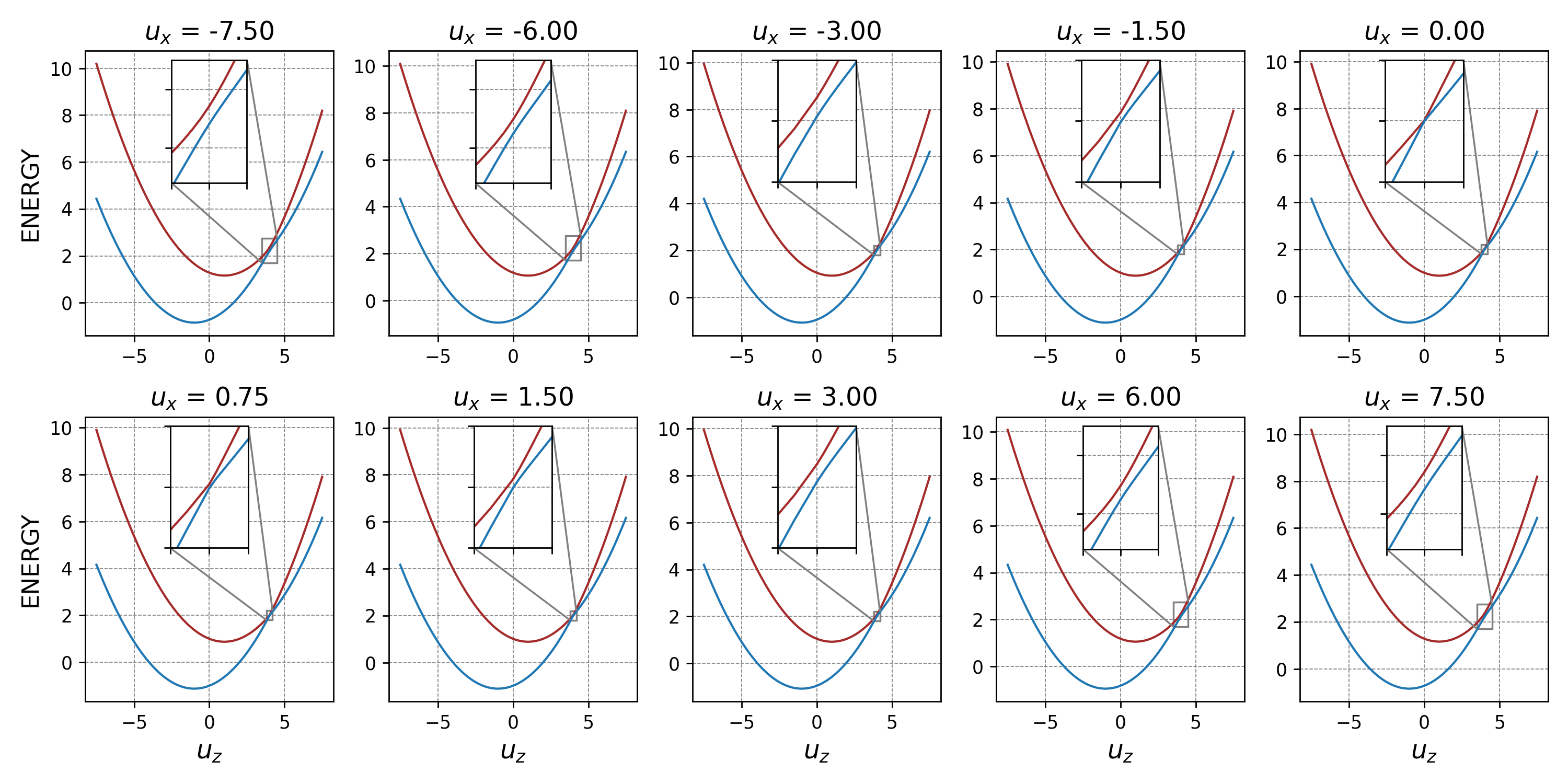} \\
    \end{tabular}
    \centering
    \caption{Potential energy surfaces over modes $u_x$ and $u_z$.
        The first left figure represents pure TET conditions
        ($E_A=E_D$, $\epsilon_x=0$ and
        $k_x^2/\omega_x^2 =k_z^2\omega_z^2 \to 1=1$).
        The last figure shows the case of away from TET conditions
        ($E_A\neq E_D$, $|E_A-E_D|>0$, $\epsilon_x \neq 0$ and
        $k_x^2/\omega_x^2 =k_z^2\omega_z^2\to 1=25$) and the projection of the PES onto the additional mode $u_x$ .}
    \label{fig:compare_pes}
\end{figure*}

In the first diagram Fig. \ref{fig:table_TET_PES} ($E_A=E_D$, $\epsilon_x=0$) the potential energy surface exhibits a characteristic symmetric conical intersection, allowing for ultrafast and efficient electron transfer between the donor and acceptor sites, resulting in a degenerate reaction path. The accompanying small plot demonstrates a linear difference between the ground and upper states along this degenerate path. This circular reaction path allows transfer in either direction around the conical intersection on the minimal PES pathway.  The same phenomenon takes place when $E_A\neq E_D$ (as shown in the middle Fig. \ref{fig:table_TET_PES}) and other TET conditions are satisfied, there is a breakdown of symmetry with a shift over the $u_z$ axis, leading to an almost linear dependence between levels and rotation angle ($\alpha$). Despite this asymmetry, there remains a conical intersection; however, electron transfer encounters greater resistance when moving to other sides of the system. The external field $\epsilon_x \neq 0$ and $E_A= E_D$, depicted in right Fig. \ref{fig:table_TET_PES}, contributes to breaking symmetry over mode $u_x$, which causes environmental deformation due to overlap integral. In this scenario, where both $E_A\neq E_D$ and $\epsilon_x\neq 0$ (middle and right Figs. \ref{fig:table_TET_PES}), electron transfer can still occur; however, the transition from donor to acceptor will not be as rapid as observed under pure TET conditions.

It is important to acknowledge that a similar challenge arises in the SSH model, where a linear approximation is utilized for the transfer integral. In the absence of damping, TET manifests as an oscillation between the Donor and the Acceptor. To induce irreversibility in TET at absolute zero temperature, non-zero damping, and a positive reaction energy $E_D-E_A$ are necessary. When there is significant disparity among the parameters in equations (\ref{tet_conditions}), the model aligns with the standard Marcus model commonly employed in chemistry to elucidate electron transfer, particularly if coupling $k_x$ is disregarded and $\epsilon_x$ remains small. Typically, a covalent coupling specified by parameter $k_x$ is not taken into account.

The most frequent scenario - normal Marcus region is illustrated in the first Fig. \ref{fig:away_tet_pes}. The greater the exergonic nature of the reaction, the smaller the obstacle for electron transfer. When both curves intersect before reaching the equilibrium position of the reactants, the activation energy rises with increasing exothermicity, which has been referred to as the Marcus inversion region (second Fig. \ref{fig:away_tet_pes}). The regime and conditions range the rate of the electron transfer. Eliminating the $u_x$ axis yields a 2D representation of Marcus' theory \cite{marcus1985electron}. For both regimes, the electron needs to overcome the barrier to be able to transfer from one side of the system (Donor) to another site (Acceptor). 

\subsection{Temperature influence}
The temperature has a gradual impact on the transfer of electrons. In pure TET, the transition from one state to another can occur at zero temperature due to the conical intersection. However, introducing white Gaussian noise as a result of temperature for both degrees of freedom ($\eta_x$ and $\eta_z$) accelerates the transfer process. Noise correlation is implemented by using a Gaussian distribution with a mean of $0$ and a variance of $\sqrt{2\gamma k_B T/\delta t}$ for both variables. To demonstrate the influence of noise, we solve the system's differential equations (Eqs. \ref{mail_set_of_equations}) employing the Runge-Kutta 5(4) method \cite{dormand1980family, shampine1986some, owren1992derivation}. The integration time step is set at $\delta t =0.001$. The strength of noise is scaled to the units of $k_BT$ where $k_B$ is Boltzmann’s constant.

To demonstrate the trajectory of the electron, we investigated two scenarios: when operating in the TET regime and when close to it ($0<E_D-E_A<=1$, $\epsilon_x=-0.001$, $k_x^2/\omega_x^2=k_z^2/\omega_z^2\rightarrow 1=1$). We aim to depict the probability evolution at both donor and acceptor sites under low temperature conditions ($k_BT=0.01$) as well as high temperature conditions ($k_BT=1.5$). The probabilities at the donor and acceptor are displayed in Fig. \ref{fig:traj_compare}. A solid line denotes the mean for each scenario, while the shaded area indicates the standard deviation. A total of 50 independent simulations were conducted with varied noise seeds to calculate these averages.

\begin{figure}[ht]
    \includegraphics[width=\columnwidth]{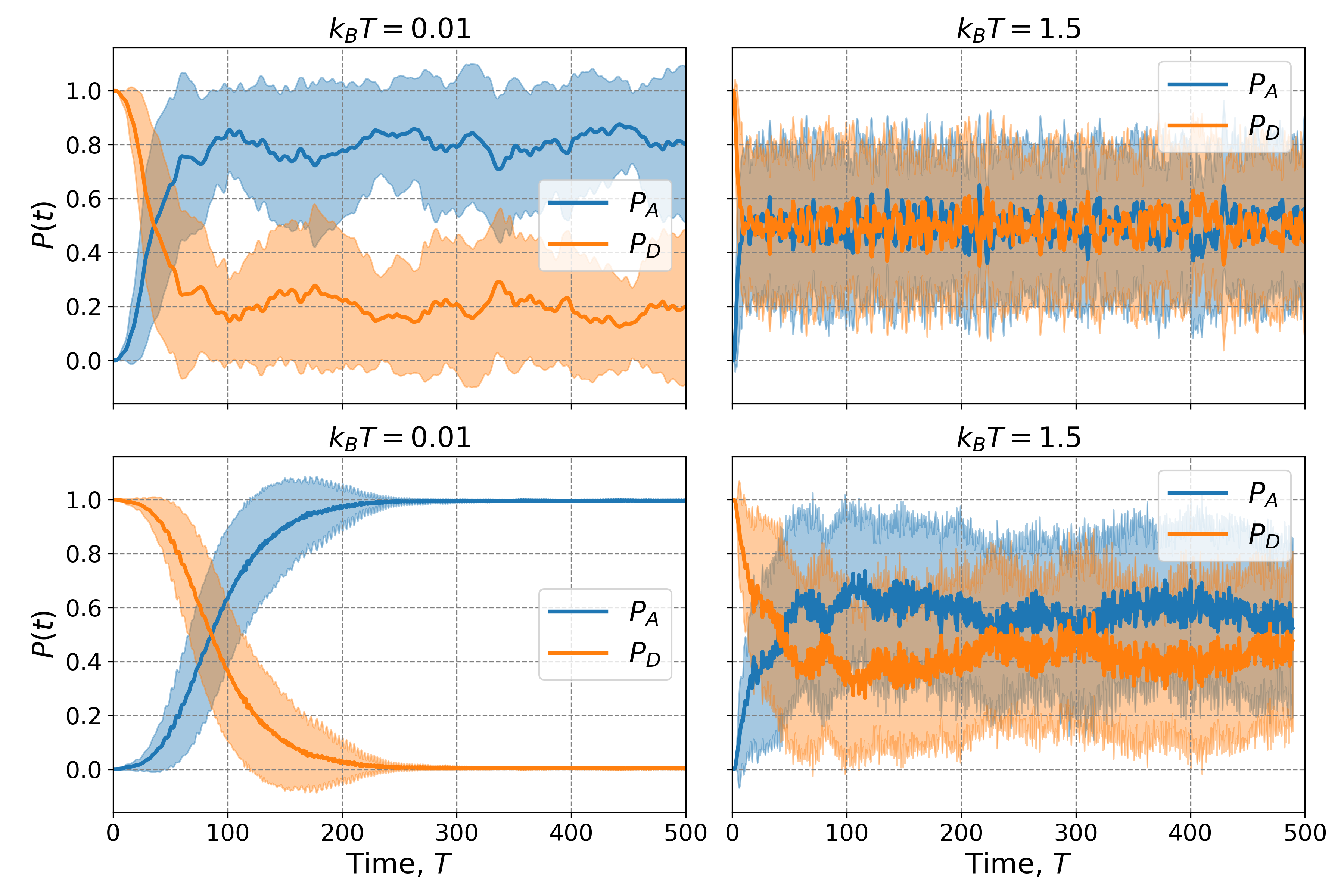} 
    \caption{The probability of donor (solid blue) and acceptor (dashed orange) over time. The thick lines indicate the mean, and the colored area represents the standard deviation. }%
    \label{fig:traj_compare}%
\end{figure}

Noise influences the evolution of probabilities over time Fig. \ref{fig:traj_compare}. In a pure TET regime, noise has a significant impact. The average value hovers around 0.8 for a temperature of $k_BT=0.01$, indicating transitions back and forth between different sites at various times. When noise levels increase, there is an observed higher frequency of transfer from donor to acceptor sites with more oscillations in the system, resulting in an average of 0.5-0.6 and a high standard deviation. The system exhibits characteristics similar to TET but with some equilibrium conditions being disrupted as portrayed in the PES (refer to middle Fig.\ref{fig:table_TET_PES}). This configuration maintains greater stability where the transfer occurs only once and minor fluctuations are visible for the $k_=0.1$ temperature range. With a noise level of $k_BT =1.l5$, we observe increased transfers and trajectories averaging around 0.S with a high standard deviation greater than in the pure TET case.

If we compare two scenarios involving pure TET with those that are distant from TET, Fig. \ref{fig:compare_pes} illustrates the PES and the projection of the PES onto the additional mode $u_x$, which is not included in Marcus's theory. In cases satisfying the TET conditions, conical intersection and degeneracy are preserved in the reaction path. Conversely, breaking the equality condition for parameters ($k_x$, $k_z$, $\omega_x$, and $\omega_z$) and introducing significant inequality leads to a visualization consistent with Marcus's theory. The intersection of PES occurs only when $u_x$ equals zero; for non-zero values, a gap exists for every other value. 

\begin{figure}[ht]
    \includegraphics[width=\columnwidth]{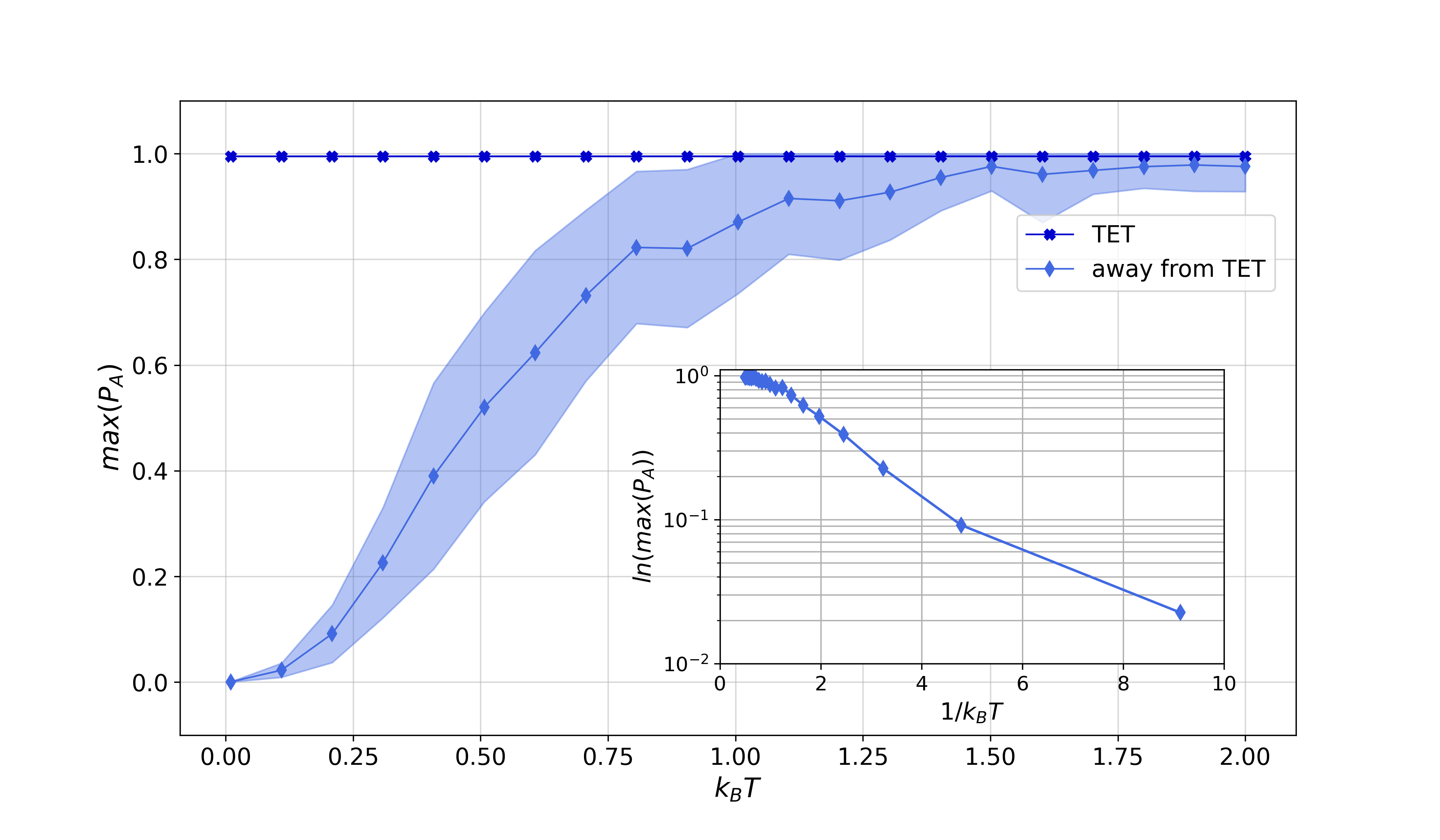} \\%
    \includegraphics[width=\columnwidth]{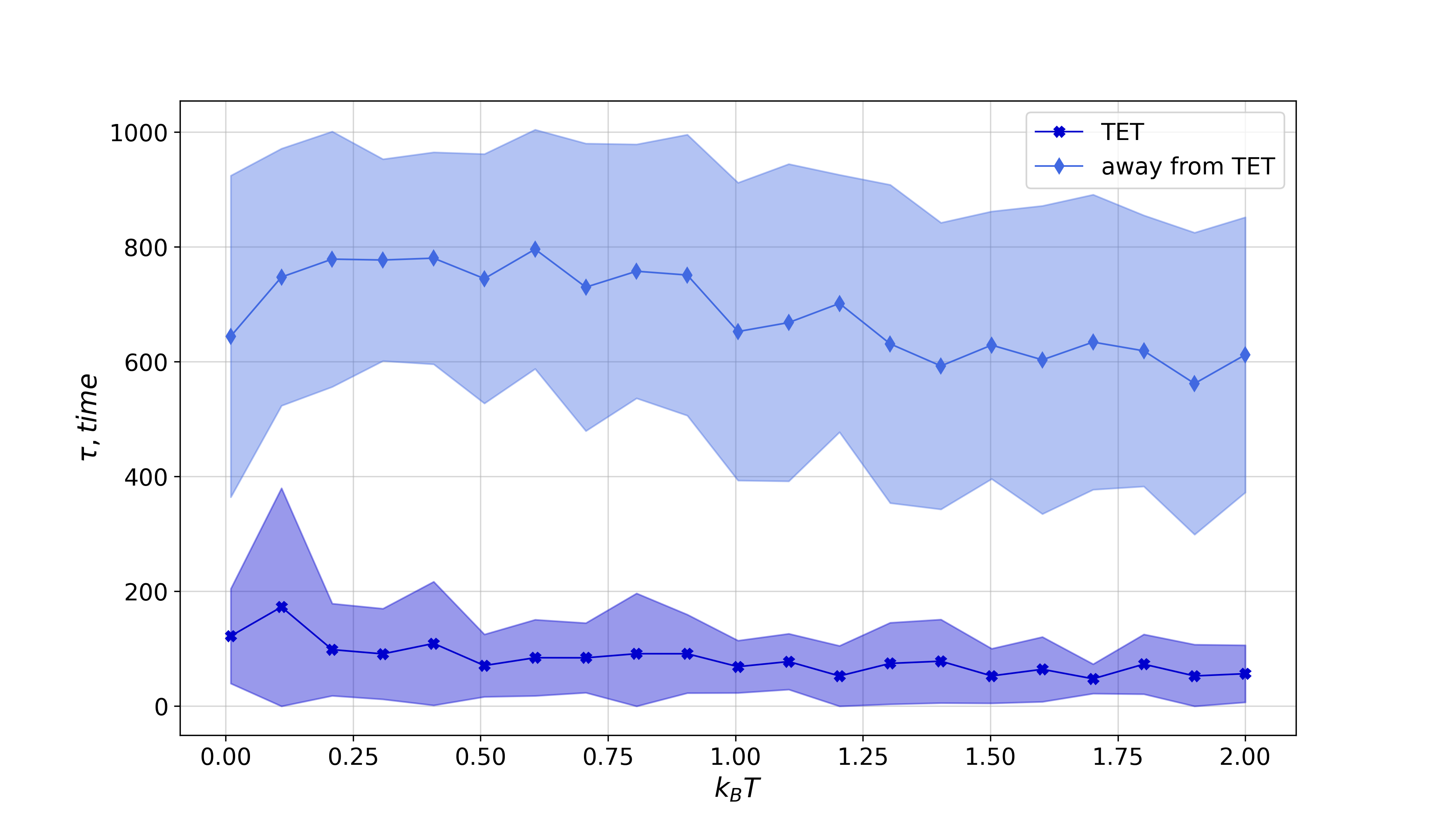}
    \caption{The mean and standard deviation of the maximum probability
        $P_A(t)$ and transfer time at the
        Acceptor sites over noise level ($k_BT$). The subplot represents the Arrhenius law for the 
        case away of the TET conditions}%
    \label{fig:comp_overtem}%
\end{figure}

To demonstrate the model's capability in analyzing the TET and Marcus theory, we compared two scenarios where the conditions described by Eqs. \ref{tet_conditions} are satisfied and where they are not. The PES shown in Fig. \ref{fig:compare_pes}, the pure TET regime, and the inverted regime of the Marcus theory.

The Arrhenius equation is another important result of chemical kinetics. According to this principle, the rate constant of electron transfer ($k_{ET}$) reduces as temperature decreases and it establishes a linear relationship between the natural logarithm of the reaction rate $ln(k_{ET})$ and the reciprocal of temperature $1/k_BT$ \cite{waskasi2016marcus}.
\begin{align}
    k_{ET} = A_{ET} \exp^{\frac{E_a}{k_B T}}
\end{align}\label{arrhe_law}
where $A_{ET}$ and $E_a$ represent the pre-exponential factor and activation energy, respectively.

Fig. \ref{fig:comp_overtem} demonstrates the average maximum probability of the Acceptor site across different temperatures for TET (navy line) and away from TET (blue line). The lower graph displays the transfer time for the maximum probability shown in the upper graph. These results were obtained from 70 runs at a specific noise level, with colored areas indicating the standard deviation. The duration over which we analyzed the transition was ten times longer than the period taken by the transition in the TET system. It is evident that in the case of TET, there is a transition between sites at every temperature value. The mean value stopped at 0.8 for the temperature 0.01 which shows that the transition for the different random seeds of noise takes place at the same region but with a slight shift. The particle is capable of oscillating in a back-and-forth manner throughout the specified interval. In contrast, when away from TET, complete transfer occurs only at a temperature equal to 1.5. A smaller diagram represents the fulfillment of Arrhenius law Eq. (\ref{arrhe_law}). The maximal probability at the Acceptor site can be understood as representing the activation energy of the reaction. It is clear to see the linear dependency on inverted temperature. 

The transition from the donor to the acceptor happens rapidly under TET conditions, as illustrated in the lower graph showing transfer time. Transfer can always be observed depending on the parameters, although the transfer time varies for each system. While temperature does not significantly increase electron transfer, it does promote oscillation across system sites. In alternative scenarios located far from TET conditions, temperature serves as a parameter driving transfer and reduces transfer time. However, there is a much larger standard deviation compared to the TET regime. The rate of transfer is four times faster in the system under TET conditions than in those situated away from such conditions.

\section{Conclusions}

Chemical reaction processes are ubiquitous and fundamental for life. The theory of chemical reactions is formulated in terms of adiabatic energy surfaces that are useful both for visualization but also quantitative understanding. In this work, we first summarized the basic theory of chemical reactions and explored some of their intricacies. The Markus theory provides a basic tool in addressing chemical dynamics through Born-Oppenheimer adiabatic surfaces. 

We focused on the TET model that utilizes non-linear resonances and leads to efficient energy and/or electron transfer in a simple donor-acceptor molecular system. This model was originally introduced as nonlinear system that has desired-engineered-transfer properties. In the present work, we show that this resonant transfer idea can be readily extended to cases where the electronic degrees of freedom are coupled additionally to vibrational ones. In this context the TET model becomes a model for special types of chemical reactions.  

Starting from first principles used in quantum chemistry we have proven the possibility of a new kind of chemical reaction called TET; the latter are coherent, ultra fast at the usual scale of chemical reactions, and still persist at moderate temperature under thermal fluctuations. This new type of chemical reactions do not obey the standard Arrhenius law. 
The TET's concern situations of chemical reactions with moderate reaction energy and thus release little energy which is the general situation in biochemistry. Such a TET chemical reaction may be found near the border of the parameter region which operates between ionic and covalent reactions. The TET occurs when ionic and covalent interactions well balance one with the other. When the TET conditions are fulfilled, there is a conical intersection between two PES which has almost cylindrical symmetry so that there is a rather flat grove connecting the initial and final state of the chemical reaction and corresponding to the reaction path.
Reversely assuming that such a conical intersection exists within a Born-Oppenheimer representation, an expansion at the lowest significant order around the conical point of intersection allows us to recover our simple model directly without using any diabatic representation, which remains physically more intuitive.
Indeed though conical intersections are a priori rare because of the Von Newmann-Wigner avoided crossing theorem, chemists were recently interested in these conical intersections for explaining
anomalous photo-induced decay.
The are now claims that from ab initio calculations that most complex molecules involving many nuclei exhibit many conical intersections between PES.

Having a  conical intersection is not sufficient for having TET; it is also required that this intersection to have at least roughly a cylindrical symmetry. In that case around the conical intersection, we have a quasi continuum of states (nuclei configurations) with almost the same energy at the biological scale that is with low energy barriers not exceeding the energy range of 0.3 eV. 
These conditions require a tuning of the model parameters which implies that TET should be highly sensitive to small perturbations of the environment of the reaction. The consequence is that relatively small perturbations of the environment may sharply slow down the reaction or even block it in the
biochemical temperature range. Otherwise, it may also reverse the TET process from the Acceptor to Donor. Such situations sometimes occur in biology depending on change on the environmental conditions such as pH concentrations of other chemical compounds, or the presence of specific molecules poisoning the biological system.
Thus, in general TET requires exceptional conditions in order to operate that a priori are very rare in inorganic chemistry. But also biochemistry of living beings also seems to be very rare in the universe, except of course on the earth. We believe that Darwinian processes over long periods of time have slowly selected and optimized this kind of chemical reaction for improving the efficiency of the living cells that use it.

We should also  mention the problem of mixed valence in inorganic chemistry. Mixed valence complexes contain an element that is present in more than one oxidation state. The Robin-Day classification distinguishes three classes. Class 1 consists of complexes where the oxidation state does not change over a long time. The change of oxidation occurs when the complex passes
through a thermally activated energy barrier. Such changes can be described by the Marcus theory. Class III consists of complexes where the oxidation state is a quantum combination of the two possible oxidation states and may be viewed as a covalent state. According to the principles of quantum mechanics, the life time of an oxidation state is related to the quantum tunneling time which is usually short. For the intermediate Class II, the time scale of the change of oxidation is much faster than for the ionic complex of class I but much slower than those of Class III which is at the time scale of purely electronic transitions. Then these mixed valence complexes are very labile that is they seem to exhibit a continuum of intermediate states between the two
oxidation states. We may claim that this situation is favorable for having TET between the two oxidation states.
Because TET is much faster than standard chemical reactions involving an energy barrier, it prevails over any other possible reaction affecting the Donor or the Acceptor because these are too slow.

The present analysis shows a path for further work. We considered here the simplest situation of elementary chemical reactions. We believe that in biological systems, TET is an ubiquitous chemical reaction that work coherently in living cells. The simpler model beyond the present dimer system is the trimer model which is a three state model. Let us consider a reaction Donor-Acceptor which involves a large energy barrier but with a small reaction energy that would be un-probable to occur spontaneously in the range of biological temperatures. 
It is then possible to choose appropriately the parameters of a third molecule $C$ (we call catalyst) which is designed especially to exhibit TET with D and next after it receives the electron from the Donor, exhibits again TET with the Acceptor.
When this enzyme binds with D and A and then forms a bridge, electron transfer occurs fast easily while this electron transfer would not occur directly. Such kinds of chemical reactions are easily affected and modulated by the environment. We found already such toy models which will be discussed in further works.

More generally, we believe it is possible to build intelligent networks of chemical reactions using elementary TET modules accomplishing well defined complex tasks.
We already presented some toy models in presentations but further studies are needed. In conclusion, we believe there is some empirical analogy with the theory of semiconductors where a simple TET dimer system would correspond to a simple Diode, the trimer catalytic model to a transistor while the global biochemical organizations of living cells would correspond to complex integrated circuits. Further work in this direction could show the generality of the ideas revolving around the TET mechanism.

\section{Acknowledgements}
The research project was co-funded by the Stavros Niarchos Foundation (SNF) and the Hellenic Foundation for Research and Innovation (H.F.R.I.) under the 5th Call of Science and Society Action - Always Strive for Excellence-Theodore Papazoglou (Project Number: 011496).

\bibliographystyle{elsarticle-num}
\bibliography{bibliography}

\end{document}